\documentclass[structabstract]{aa}
\usepackage{graphicx}
\usepackage{txfonts}
\usepackage{longtable}
\usepackage{supertabular}
\usepackage{lscape}
\usepackage{natbib}

\begin{document}
\title{The instrumental polarization of the Nasmyth focus polarimetric differential imager NAOS/CONICA (NACO) at the VLT}

\subtitle{Implications for time-resolved polarimetric measurements of Sgr~A*}

\author{G. Witzel\inst{1}\fnmsep\thanks{e-mail: witzel@ph1.uni-koeln.de}, A. Eckart\inst{1,2}, R. M. Buchholz\inst{1}, M. Zamaninasab\inst{2,1}, R. Lenzen\inst{3}, R. Sch\"odel\inst{4}, C. Araujo\inst{1}, N. Sabha\inst{1}, \\
M. Bremer\inst{1}, V. Karas\inst{5}, C. Straubmeier\inst{1}, \and K. Muzic\inst{6} }
\institute{I. Physikalisches Institut der Universit\"at zu K\"oln (PH1), Z\"ulpicher Stra\ss e 77, 50937 K\"oln, Germany
\and
Max-Planck-Institut f\"ur Radioastronomie (MPIfR), Auf dem H\"ugel 69, 53121 Bonn, Germany
\and
Max-Planck-Institut f\"ur Astronomie (MPIA), K\"onigstuhl 17, 69117 Heidelberg, Germany
\and
Instituto de Astrof\'{i}sica de Andaluc\'{i}a - CSIC, Glorieta de la Astronom\'{i}a S/N, 18008 Granada, Spain
\and
Astronomical Institute of the Academy of Sciences, Bocni II 1401/1a, CZ-141 31 Praha 4, Czech Republic
\and
Department of Astronomy and Astrophysics, University of Toronto, 50 St. George Street, M5S~3H4 Toronto ON, Canada
}
\date{Preprint online version: 2010 Oktober 22}
\titlerunning{The instrumental polarization of NACO}
\authorrunning{G. Witzel et al.}
\abstract
{We report on the results of calibrating and simulating the instrumental polarization properties of the 
    ESO VLT adaptive optics camera system NAOS/CONICA (NACO) in the Ks-band.}
{Our goal is to understand the influence of systematic calibration effects on the time-resolved polarimetric 
    observations of the infrared counterpart of the Galactic center super-massive black hole at the position 
    of Sagittarius~A* (Sgr~A*).}
{We use the Stokes/Mueller formalism for metallic reflections to describe the instrumental polarization. 
    The model is compared to standard-star observations and time-resolved observations of bright sources 
    in the Galactic center. The differences between calibration methods are simulated and tested for three 
    polarimetric Ks-band light curves of Sgr~A*.}
{We find the instrumental polarization to be highly dependent on the pointing position of the telescope 
    and about 4\% at maximum. Given the statistical uncertainties in the data acquisition, the systematic 
    effects of the employed calibration method are negligible at high-time resolution, as it is necessary and achieved for in the case of Sgr~A*.
    We report a polarization angle offset of $13.2^{\circ}$ due to a position angle offset of the $\lambda/2$-wave plate with respect to the header value that affects the calibration of NACO data taken before autumn 2009.}
{With the new model of the instrumental polarization of NACO it is possible to measure the polarization 
    with an accuracy of 1\% in polarization degree. The uncertainty of the polarization angle is $\le5^{\circ}$ for polarization degrees $\ge 4\%$.
    For highly sampled polarimetric time series we find that the improved understanding of the polarization properties gives results that are fully consistent
    with the previously used method to derive the polarization.
    The small difference between the derived and the previously employed polarization calibration is well within the 
    statistical uncertainties of the measurements, and for Sgr~A* they do not affect the results from 
    our relativistic modeling of the accretion process.}
\keywords{Instrumentation: polarimeters, techniques: polarimetric, Polarization, infrared: general, black hole physics, Galaxy: center}
\maketitle

\section{Introduction}

The polarization of electromagnetic radiation is an essential piece of information to determine the nature of emission processes and the physical parameters of the environment in which the radiation is generated. This is true in particular for the emission from Sagittarius A* (Sgr~A*)
at the center of the Milky Way. This source is associated with the nearest super-massive black hole candidate ($\sim 4\times10^6 M_\odot$), as inferred from motions of stars near the Galactic center (\citealt{1996Natur.383..415E,1997MNRAS.284..576E}; \citealt{2002MNRAS.331..917E}; \citealt{2002Natur.419..694S}; \citealt{2003ApJ...597L.121E}; \citealt{2000Natur.407..349G,2005ApJ...620..744G,2008ApJ...689.1044G}; \citealt{2009ApJ...692.1075G}).

Since the first near-infrared (NIR) polarimetric Wollaston prism observation of Sgr~A* 
in 2004 (\citealt{2006A&A...450..535E}), polarized flares have been regularly observed (\citealt{2006A&A...460...15M,2006A&A...458L..25M,2007A&A...473..707M}; \citealt{2008A&A...479..625E}, \citealt{2010A&A...510A...3Z}). The NIR counterpart to Sgr~A* is extremely variable with short bursts of increased emission exceeding 5 mJy, which occur four to six times a day and last typically for about 100 minutes. The linear polarization degrees can reach 20\% to 50\% of the total intensity.

The polarized flares are often associated with simultaneous X-ray flares (\citealt{2001Natur.413...45B}; \citealt{2003A&A...407L..17P,2008A&A...488..549P}; \citealt{2003Natur.425..934G}; \citealt{2004A&A...427....1E,2006A&A...450..535E,2006A&A...455....1E,2006Msngr.125....2E,2008A&A...479..625E,2008JPhCS.131a2002E,2008A&A...492..337E};
\citealt{2006A&A...460...15M,2006A&A...458L..25M,2007A&A...473..707M}; \citealt{2006ApJ...644..198Y,2006ApJ...650..189Y,2007ApJ...668L..47Y,2008ApJ...682..361Y}; \citealt{2009ApJ...698..676D}; \citealt{2010A&A...512A...2S}).
This strongly suggests synchrotron-self-Compton (SSC) or inverse Compton emission as the
responsible radiation mechanism (\citealt{2004A&A...427....1E,2006A&A...450..535E,2006A&A...455....1E}; \citealt{2004ApJ...606..894Y}; \citealt{2006ApJ...648.1020L}).

Some models that have been applied successfully to the observations assume the flare phenomenon to be linked to emission from single or multiple hot spots near the last stable orbit of the black hole. The characteristic behavior of general relativistic flux modulations that are produced via such orbiting hot spots have been discussed earlier
(see e.g. \citealt{1973ApJ...183..237C}; \citealt{1977Natur.266..429S}; \citealt{1991A&A...245..454A}; \citealt{1992A&A...257..531K}; \citealt{1995ApJ...448L..21H}; \citealt{2004ApJS..153..205D,2008MNRAS.384..361D}, \citealt{2010A&A...510A...3Z}).

Based on relativistic models \cite{2010A&A...510A...3Z} find a correlation between the modulations of the observed flux density light curves and changes in polarimetric data. The authors also confirm that this correlation is predicted by the hot spot model. Correlations between intensity and polarimetric parameters of the observed light curves and a comparison of predicted and observed light curve features through a pattern recognition algorithm result in the detection of a signature possibly associated with orbiting matter under the influence of strong gravity. This pattern is found to be statistically significant against randomly polarized red noise.

The investigation of the emission from Sgr~A* and the application of the model calculations to the observed light curves of polarized light decisively depend on the quality of the polarization calibration. The crucial NIR polarization data were obtained through NACO, a differential polarimetric imager, at the ESO VLT UT4. This system is mounted at a Nasmyth focus of the altitude azimuth mounted UT4 telescope, which complicates the exact polarimetric calibration.
Therefore we carried out a detailed analysis of the instrumental properties of this system, determined the systematic instrumental uncertainties and discuss their influence on the Sgr~A* measurements and the consequences for the astrophysical interpretation of the light curves obtained in polarized light.

The goal of this paper is to investigate the instrumental polarization (IP) mainly on a base of scientific data, which are the outcome of eight years of observations of Sgr~A*. These data are optimized for the astrophysical time series analysis of Sgr~A* with high-time resolution. They nevertheless provide enough information to tackle two aspects of the systematic instrumental effects: On one hand the description of the IP and its behavior in absolute values with an accuracy of about 1\% in linear polarization degree, and on the other hand the systematic uncertainties of the time variability of the polarimetric parameters of Sgr~A*.

The polarimetric mode of NACO considered here does not provide information on circular polarization, and the flat-field calibration data are not optimized for polarimetric measurements. Both facts necessitate complicated procedures for calibrating the existing data. We describe these procedures here in detail. We also give ideas on how to improve the polarimetric calibration for future observations that do not require a high-time sampling, but aim to measure polarized emission to an accuracy of even a few tenth of a percent in linear polarization degree.

In section~\ref{instpol} we give a detailed description of the instrumental polarization of NACO in Ks-band and a correction algorithm using a model based on material constants of standard coatings. Such a model has not yet been available to the comunity for this telescope/camera combination and therefore could not be applied to observations of Sgr~A* until now. 

In section~\ref{comp} we compare the model with observations of standard stars and time-resolved polarimetric measurements of bright stars in the Galactic center and analyze to what degree we can correct the systematics. 

In section~\ref{cacom} commonly used calibration procedures are compared with the new calibration method. Systematic differences between the methods and their influence on the variability measurements of the polarization degree and angle of Sgr~A* are discussed. 

In section~\ref{concl} we summarize our results and their impact on the interpretation of polarimetric time series of Sgr~A*.

\section{A model for the instrumental polarization of NACO \label{instpol}}
NACO is an adaptive optics imager for the NIR at the Very Large Telescope (VLT) that is run by ESO. For a detailed description of NACO see \cite{2003SPIE.4841..944L}, \cite{2003SPIE.4839..140R}, and \cite{nacomanual}. It provides a mode for polarimetric differential imaging combining a Wollaston prism (in the following referred to as Wollaston), which allows for measuring two orthogonal angles simultaneously, and a $\lambda/2$ wave plate (HWP). In observation periods before 2008 a wire grid mode was available as well. To determine the instrumental polarization (IP) of this instrument we use the Stokes and Mueller calculus. It allows us to describe the influence of optical elements on the polarization. The model presented in this section enables us to determine the IP as a function of the parallactic angle. It is developed following a model for the IP of the Telescopio Nazionale Galileo (TNG) presented in \cite{2003SPIE.4843..456G}. Since it is crucial for the following to have a clear definition of conventions and variables we first introduce some basic formalism.

\subsection{Polarimetry with NACO: basics and conventions}

The projection of the electric field vector $E$ of fully linearly polarized light onto a preferential direction that makes an angle $\psi$ with the E-vector is given by
\begin{equation}
E_{\psi} = E_{0} \cos(\psi) \;.
\end{equation}
The energy carried by the projected electromagnetic wave and thus the intensity is proportional to $E^{2}$:
\begin{equation} \label{eq2}
I \sim E_{\psi}^{2} = E_{0}^{2}\cos^{2}(\psi) \;.
\end{equation}
For partially linearly polarized light - defined by $I_{tot}$ the total intensity, $P$ the degree of linear polarization, and $\phi$ the polarization angle - the dependency of the intensity on the angle $\theta$ of the preferential direction is given by
\begin{equation} \label{chan}
I_{P,\phi}(\theta) = \frac{I_{tot}}{2} + \frac{PI_{tot}}{2} \cos(2 \left[\theta-\phi\right]) = I(\theta) \;.
\end{equation}
$\theta$ is measured with respect to the polarization angle reference that defines $\phi$ and commonly is given by the north-south axis on the sky\footnote{ It is the direction of the linear polarization if $U=0$ (see Eq.~\ref{Qdef}) and also defines the orientation of the Mueller matrices below.}. $I(\theta)$ is the intensity we would measure with an analyzer at the angle position $\theta$ and a transmittance of unity. Please see Fig.~\ref{vect}.

A convenient tool to describe partial polarization of incoherent light is the Stokes formalism. The normalized Stokes vector for partial (linear) polarization is defined as
\begin{equation}
S =
\left(\begin{array}{c}
I_{tot} = 1 \\
Q \\
U \\
V \hspace{0.1cm}
\end{array}\right) \;,
\end{equation}
with
\begin{eqnarray} \label{Qdef}
Q & = & \frac{I(0^{\circ})-I(90^{\circ})}{I(0^{\circ})+I(90^{\circ})} \nonumber \\
U & = & \frac{I(135^{\circ})-I(45^{\circ})}{I(135^{\circ})+I(45^{\circ})} \;.
\end{eqnarray}
$Q$ and $U$ represent the linear polarization and $V$ the circular polarization (NACO does not provide a $\lambda/4$-wave plate to measure V). The parameters $P$ and $\phi$ are related to $Q$ and $U$ by
\begin{eqnarray}
P & = & \sqrt{Q^2 + U^2} \nonumber \\
\phi & = &  \frac{1}{2}\arctan\left(\frac{U}{Q}\right) \;.
\end{eqnarray}

In this formalism the influence of any optical element on the intensity and polarization can be expressed by a linear operation on the Stokes vector:
\begin{equation}
S' = M \times S \;,
\end{equation}
where $M$ is the Mueller matrix of the element, S the Stokes vector of the incoming light and S' the Stokes vector of the outgoing light. The elements of the Mueller matrix represent the linear dependency of each Stokes parameter in $S'$ on those in $S$.

\begin{figure}
   \centering
   \includegraphics[angle=-90,width=8.5cm]{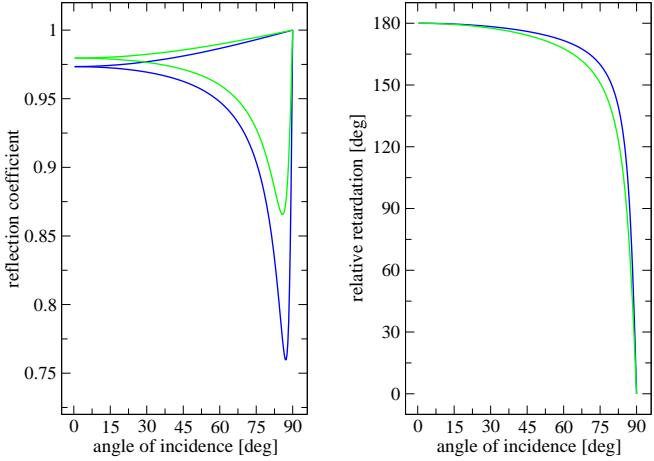}
      \caption{Reflection properties of metallic surfaces in Ks-band. The green curves represent gold, the blue aluminum.}
         \label{rho}
\end{figure}

The optical elements considered here are mainly mirrors with metallic coatings. For these surfaces every incident electromagnetic wave can be decomposed in a component parallel to the plane of incidence and one perpendicular to this plane. As described in detail by \cite{2003SPIE.4843..456G} and \cite{1973A&A....24..165C}, reflections at metallic surfaces have two effects: (1) a reflection introduces a linear polarization because the reflectivity for these components is different; (2) the reflection causes a circular polarized contribution by shifting the phase between the components. Both effects can be described by a Mueller matrix that combines the matrix elements for a linear polarizer and a retarder:
\begin{equation}
\label{R}
R = \left(
\begin{array}{cccc}
\frac{1}{2}(r_{\perp} + r_{\parallel}) & \frac{1}{2}(r_{\perp} - r_{\parallel}) & 0 & 0 \\
\frac{1}{2}(r_{\perp} - r_{\parallel}) & \frac{1}{2}(r_{\perp} + r_{\parallel}) & 0 & 0 \\
0 & 0 & \sqrt{r_{\perp}r_{\parallel}}\cos(\delta) & -\sqrt{r_{\perp}r_{\parallel}}\sin(\delta) \\
0 & 0 & \sqrt{r_{\perp}r_{\parallel}}\sin(\delta) & \sqrt{r_{\perp}r_{\parallel}}\cos(\delta)
\end{array}
\right) \;,
\end{equation}
with $r_{\perp}$ and $r_{\parallel}$ the reflection coefficients for the two components and $\delta$ 
the relative retardation between the components. With the material-dependent refractive index and 
extinction coefficient all three parameters can be calculated by using the Fresnel formulae.
In  the left plot of Fig.~\ref{rho} we show the reflection coefficients for p- and s-waves as a 
function of the angle of incidence for Ks-band. In the right plot of Fig.~\ref{rho} we  show the relative retardation 
a mirror hit by a perpendicular beam is considered to have the same effect as a HWP.
The material constants (Ks-band) used for these plots are listed in Table~\ref{mat}.

Eq.~\ref{R} defines the Mueller matrix for metallic reflection in a way that the preferential direction of the matrix (the direction of the introduced polarization, always perpendicular to the plane of incidence) is oriented parallel to the polarization angle reference (generally north-south). To change the (perpendicular) orientation of the plane of incidence with respect to the polarization angle reference by an angle $\gamma$, one has to apply the transformation
\begin{equation} \label{trans}
R' = T(-\gamma) \times R \times T(\gamma) \;,
\end{equation}
with T the rotation in Stokes space, which is defined as
\begin{equation} \label{stokesrot}
T(p) = \left(
\begin{array}{cccc}
1 & 0 & 0 & 0 \\
0 & \cos(2p) & \sin(2p) & 0 \\
0 & -\sin(2p) & \cos(2p) & 0 \\
0 & 0 & 0 & 1
\end{array}
\right) \;.
\end{equation}

\begin{figure}
   \centering
   \includegraphics[angle=0,width=8cm]{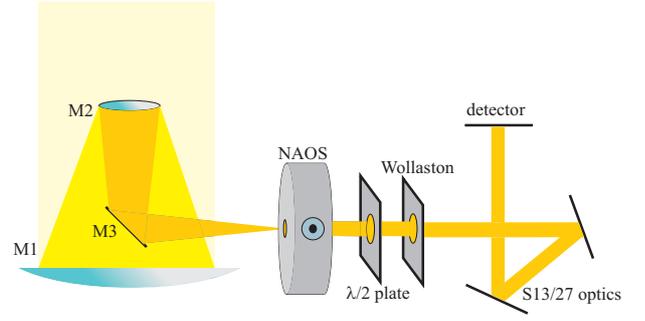}
      \caption{Optical elements of UT4, NAOS \& CONICA and their relative orientation in the moment of the meridian transit. 
              }
         \label{train}
\end{figure}

NACO is mounted at the Nasmyth focus of VLT Yepun. Because of the $45^{\circ}$ tilted folding mirror M3 (see Fig.~\ref{train}) that sends out the beam to the Nasmyth focus, NACO has a significant instrumental polarization that is depending on the parallactic angle. In the ESO user manual the total IP is estimated to be up to 4\%. NACO provides various setups for polarimetry combining different cameras (with different pixel scales) and filters with the Wollaston and the HWP. A table of available cameras and filters is shown in \cite{nacomanual}. Here we are concentrating on the most commonly used setup: The cameras S13 and S27 with the Ks-band filter, the Wollaston, and the HWP.

\begin{figure}
   \centering
   \includegraphics[width=8cm]{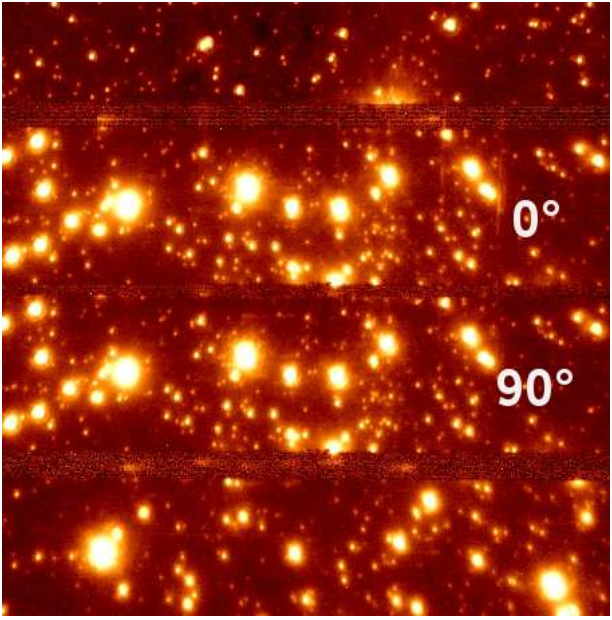}
      \caption{Polarimetric observations with the Wollaston: the picture shows a frame observed with the S13 camera.
               The image taken with the ordinary beam of the Wollaston is the upper stripe.
              }
         \label{fov}
\end{figure}

\begin{figure}
   \centering
   \includegraphics[angle=0,width=6cm]{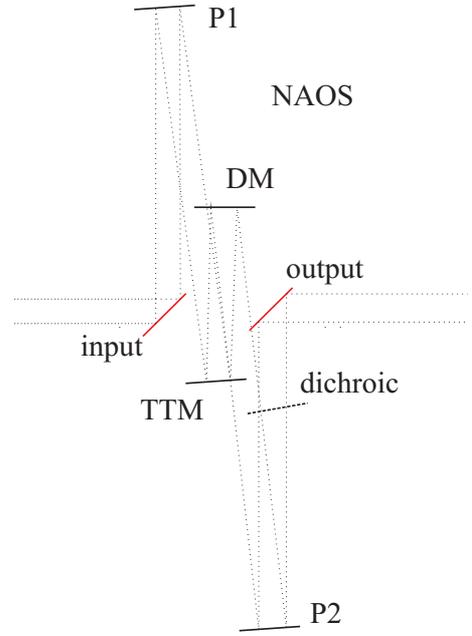}
      \caption{Light train in the adaptive optics module NAOS
              }
         \label{NAOS}
\end{figure}

The field of view of S13 in combination with the Wollaston analyzer and the polarimetric mask is shown in Fig.~\ref{fov}.
The upper stripe is the ordinary beam ($0^{\circ}$) and the the lower one the extraordinary ($90^{\circ}$). The angle positions of the HWP in the header\footnote{The angle position of the HWP is reported in the NACO FITS header under the keyword "INS RETA2 ROT"; the encoder position can be found under "INS ADC1 ENC".} of the NACO data is counted with the same sense of rotation. Note that this sense is reverse to the sky because the number of mirrors is odd and every mirror turns it once (see Fig.~\ref{train} and Fig.~\ref{NAOS}).

Not only M3 contributes to the IP of NACO. Every significantly inclined reflective surface in the light train has to be included in a model of polarimetric instrumental systematics. All optical elements (including the analyzer) and their relative orientations are discussed below.
In Fig.~\ref{train} we show the optical elements of UT4 and NACO. In this sketch the direction of the optical train within NAOS (defined by the connection line between the input mirror and P1 in Fig.~\ref{NAOS}) is perpendicular to the paper plane and indicated by the face-on arrow sign on NAOS. A sketch of NAOS is shown in more detail in Fig.~\ref{NAOS}.

A de-rotation of the parallactic angle rotates the optical 
train beginning with NAOS with respect to the mirrors M1/2/3. The elevation rotation is 
only affecting the hole assembly because the rotator of NACO compensates for it. It does not influence the IP.
After HWP and Wollaston there are two more folding 
mirrors in the architecture of CONICA; the inclinations of these two mirrors are different 
for different cameras (see Table~\ref{mat} and angles of incidence $\epsilon$ therein).
The light train of NAOS in Fig.~\ref{NAOS} shows that there are only two mirrors with a significant 
inclination: the input and output mirror (red). The parabolic mirrors P1 and P2, 
the tip-and-tilt mirror TTM, and the deformable DM have inclinations $\le5^{\circ}$ and 
can be neglected. The specifications of the dichroic are not accessible to us, but here the 
inclination of only about $12^{\circ}$ is also very small.

\subsection{Instrumental polarization generated by M3}

M3 is coated with aluminum and inclined by $45^{\circ}$. To understand the position-dependent part of the IP, it is important to analyze the time-depending orientation of M3.

The common angle reference at the sky is the north-south axis. At the Nasmyth focus this direction is parallel to the plane defined by the edge of the main mirror M1 in the moment of the meridian transit of the source. In this moment the north-south axis is parallel to the preferential direction of the matrix in Eq.~\ref{R} (see Fig.~\ref{train}). At all other moments it is tilted by $-p$ with respect to the preferential direction of matrix~(\ref{R}) (with $p$ the parallactic angle).

In order to transform matrix~(\ref{R}) to the north-south reference at any moment, we have to use a transform similar to Eq.~\ref{trans}. After changing the reference from the celestial system to the reference system of M3 and applying R for the metallic reflection, we have to transform to the detector reference. The detector is de-rotated, it follows the elevation rotation of the Nasmyth focus and the parallactic rotation of the source\footnote{i.e. the angle between main mirror and detector orientation only depends on the parallactic rotation.}. After the reflection this parallactic rotation has the reverse sense, thus the position depending part of the IP can be expressed as

\begin{equation}
M_{M3} = T(p) \times R_{alu} \times T(p) \;,
\end{equation}
where $R_{alu}$ is the reflection matrix for bare aluminum with the values listed in Table~\ref{mat}.

\subsection{A description of the entire instrumental polarization of NACO}
\label{totip}

\subsubsection*{NAOS}

The two $45^{\circ}$ inclined Silflex coated folding mirrors in the adaptive optics module NAOS are described by the square of matrix~(\ref{R}):
\begin{equation}
M_{NAOS} = R_{sil} \times R_{sil} \;.
\end{equation}
The material constants for Silflex have been provided by the producer Balzers Optics and can be found in Table~\ref{mat}.

\subsubsection*{The HWP}

\begin{figure}[t!]
   \centering
   \includegraphics[angle=-90,width=8cm]{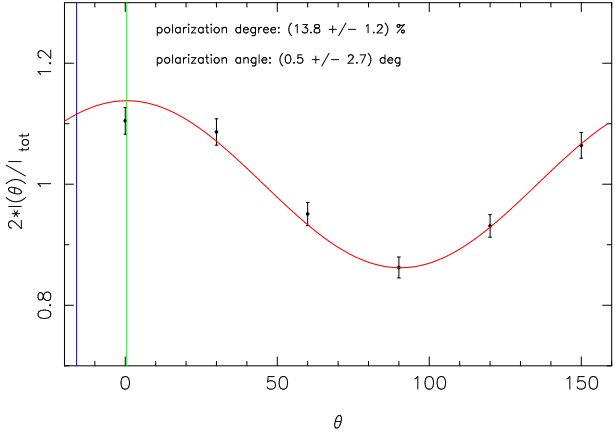}
      \caption{IRS21 in 2005 (Wollaston). The plot shows the intensity divided by $I_{tot}/2$ as a function of angle $\theta$ (as defined in Eq.~\ref{chan}). The data are obtained from mosaics. No calibration was applied. The green line marks the measured polarization angle, the blue the published angle ($9.8 \%$@$14^{\circ}$) from \cite{1999ApJ...523..248O}. In this plot the angle is counted in the instrument sense (negative with respect to sky).}
         \label{irs21off}
\end{figure}

\begin{table}
      \caption[]{Encoder positions of the HWP.}
         \label{enc}
     $$
         \begin{array}{ll}
            \hline
            \hline
            \noalign{\smallskip}
            \rm{Encoder\: steps} & \rm{Angle} \\
            & [\rm{deg}] \\
            \noalign{\smallskip}
            \hline
            \noalign{\smallskip}
            \rm{Manual:} & \\
            x &  \alpha = (x+205)*0.08789 \\
            -205 \sim 3891 & 0 \\
            51 & 22.5^{\circ} \\
            0 & 205*0.08789 = 18.02 \\
            \noalign{\smallskip}
            \hline
            \noalign{\smallskip}
            \rm{Revision:} & \\
            0 & (11.4 \pm 0.2) \\
            \noalign{\smallskip}
            \hline
            \noalign{\smallskip}
            \rm{Difference:} & \\
            & (-6.6 \pm 0.2)  \\
            \noalign{\smallskip}
            \hline
          \end{array}
     $$
\tablefoot{The table shows that the actual reference system for the HWP is offset by $(-6.6 \pm 0.2)^{\circ}$ with respect to the reference system assumed in the manual. This results in a positive angle offset of $13.2^{\circ}$ for the polarization channels (in the instrumental sense of rotation). One turn of the plate corresponds to 4096 encoder steps. An angle of $0^{\circ}$ actually corresponds to an encoder position of 3966.}
   \end{table}

The polarimetric analyzer is part of the camera CONICA. It mainly consists of the HWP, the Wollaston, and the detector. The HWP turns the angle of the linearly polarized part of the light by the double of its position angle. The position angle is the angle between the fast axis (or slow axis, the degeneracy is $90^{\circ}$) and the polarimetric angle reference. The formula (see Table~\ref{enc}) given in the ESO manual 
for the dependency of the position angle (angle with respect to the north-south axis) on the encoder steps has to be modified. The plot in Fig.~\ref{irs21off} shows the un-calibrated polarization of IRS21 in a dataset taken in 2005 with the Wollaston. The dataset exhibits an offset in polarization angle of about $14^{\circ}$ in comparison with \cite{1999ApJ...523..248O}. A maintenance of NACO in autumn 2009 revealed the actual position angle reference. The true offset is $(13.2 \pm 0.3)^{\circ}$ (see Table~\ref{enc}), a value that agrees very well with the value of $14^{\circ}$ that we predicted from the observational data before the intervention. The previously reported angular offset of $34^{\circ}$ (\citealt{2007MNRAS.375..764T}) could not be verified. This value may result from the offset we found combined with the fact that \cite{2007MNRAS.375..764T} use a sine- rather than a cosine-function to describe the 2$\theta$-dependency of the polarized channel flux. This causes a $45^{\circ}$-shift in the opposite direction to the HWP offset, resulting in a total of $32^{\circ}$. 

The HWP was installed on 2003 August 8, and we assume that the offset is constant for all epochs since then.
\\

\subsubsection*{Total IP}

Because the plane of incidence of the NAOS light train is perpendicular to the plane of incidence of M3, we have to transform $M_{NAOS}$ according to Eq.~\ref{trans} with $\gamma=90^{\circ}$. We also have to add a rotation matrix for the rotator adapter that de-rotates the instrument and can change the orientation of the field of view (from north-south on the y-axis of the detector to any angle). The discussed matrices and orientations finally result in a total Mueller matrix:
\begin{eqnarray} \label{totipmat}
M_{NACO} & = & (Tr \; \times) \; T(-\beta) \times T(90^{\circ}) \times M_{NAOS} \nonumber \\
 & & \times T(-90^{\circ}) \times T(\alpha) \times M_{M3} \nonumber \\
 & = & (Tr \; \times) \; T(-\beta) \times T(90^{\circ}) \times R_{sil} \times R_{sil} \times T(-90^{\circ}) \nonumber \\
 & & \times T(\alpha) \times T(p) \times R_{alu} \times T(p) \;,
\end{eqnarray}
with $-\alpha$ the angle of rotator adapter as reported in the ESO FITS header keyword "ADA POSANG" and $ \beta = 13.2^{\circ}$ the offset of the HWP. The matrix $Tr$ represents the effects of the analyzer and its transmission. It is included here for the sake of completeness. A more detailed discussion of these effects and their correction will follow in section~\ref{polanaly}.
\\

\subsection{The instrumental polarization in numbers}

In this section we investigate the behavior of the introduced model. All material-dependent parameters of this model are summarized in Table~\ref{mat}. These parameters are mainly literature values for the materials. In section~\ref{comp} the model is compared with standard stars and light curves of bright GC stars, which exhibit the variations of the IP with the parallactic angle. To match the observations the material constants $k$ for aluminum and $\delta$ for the Silflex coating had to be slightly changed as described in the captions of Table~\ref{mat}. Here we already discuss the final version of the model that is gauged with the calibration sources. In section~\ref{comp} we will then justify the model and the chosen parameters.

\begin{table}
      \caption[]{Material constants of coatings.}
         \label{mat}
     $$
         \begin{array}{llcccccc}
            \hline
            \hline
            \noalign{\smallskip}
            \rm{Coating} & \rm{Mirror} &  n  & k & \epsilon & r_{\parallel} & r_{perp} & \delta \\
            & & & & [\rm{deg}] &  & & [\rm{deg}] \\
            \noalign{\smallskip}
            \hline
            \noalign{\smallskip}
            \rm{Alu.} & \rm{M3} & 2.75  & 20.0 & 45^{\circ} & 0.96262 & 0.98113 & 176.03 \\
            \rm{Silflex} & \rm{NAOS \hspace{0.1cm} I \& II} &    -    &  -  & 45.00 & 0.98272 & 0.98872 & 165.00 \\
            \rm{Gold} &  \rm{S13 \hspace{0.1cm} I} & 0.99     &   13.8 & 42.05 & 0.97266 & 0.98484 & 175.02\\
             &       \rm{S13 \hspace{0.1cm} II} & &    & 2.95 & 0.97959 & 0.97964 & 179.98 \\
             & \rm{S27 \hspace{0.1cm} I} &        &    & 27.65 & 0.97701 & 0.98193 & 177.99 \\
             & \rm{S27 \hspace{0.1cm} II} &       &    & 17.35 & 0.97865 & 0.98054 & 179.23 \\
            \noalign{\smallskip}
            \hline
          \end{array}
     $$
\tablefoot{Refractive index $n$, extinction coefficient $k$, reflection coefficients, and relative retardation for the coatings of NACO at $2.2\mu$. $\epsilon$ is the angle of incidence for which the reflection coefficients and relative retardation was computed. The material constants can be found on www.RefractiveIndex.info. The default $k$ value for aluminum is 22.3, and $\delta$ for Silflex is 166.6 according to Balzers Optics specifications. Both values have been changed to match the position dependency of measured Stokes parameters (see section~\ref{gaug}).}
   \end{table}

We are now able to evaluate the contributions of the different optical elements quantitatively. First we numerically express the matrices of Eq.~\ref{totipmat}. For the given material parameters and a parallactic angle of $0^{\circ}$ we find $M_{M3}$ to be
\begin{equation}
M_{M3} = \left(
\begin{array}{cccc}
0.972 & 0.009 & 0 & 0 \\
0.009 & 0.972 & 0 & 0 \\
0 & 0 & -0.969 & -0.067 \\
0 & 0 & 0.067 & -0.969
\end{array}
\right) \;,
\end{equation}
and for a parallactic angle of $45^{\circ}$
\begin{equation}
M_{M3} = \left(
\begin{array}{cccc}
0.972 & 0 & 0.009 & 0 \\
0 & 0.969 & 0 & -0.067 \\
-0.009 & 0 & -0.972 & 0 \\
0 & -0.067 & 0 & -0.969
\end{array}
\right) \;.
\end{equation}
Obviously the main effect of the tertiary mirror is an $I \rightarrow Q/U$\footnote{Depending on the parallactic angle, this cross talk is affecting $Q$ ($p=0^{\circ}/90^{\circ}$), $U$ ($p=45^{\circ}/135^{\circ}$), or both of them.} cross talk of about $1\%$ of the total intensity. The cross talks between linear and circular polarization can be on the order of $7\%$ of $Q/U$ or $V$ respectively. For weakly polarized sources this is a minor effect. The transformed $M_{NAOS}$ is given by
\small
\begin{equation}
T(90^{\circ}) \times M_{NAOS} \times T(-90^{\circ}) = \left(
\begin{array}{cccc}
0.972 & -0.006 & 0 & 0 \\
-0.006 & 0.972 & 0 & 0 \\
0 & 0 & 0.841 & -0.486 \\
0 & 0 & 0.486 & 0.841
\end{array}
\right) \;.
\end{equation}
\normalsize
Here $Q$ is affected by about $0.6\%$ of $I$ and the $U \leftrightarrow V$ cross talks are on the order of $50\%$ of the corresponding values $U$ or $V$ respectively. With these numerical matrices, Eq.~\ref{totipmat}, and the parameter $\alpha$ set to zero, we can write matrix $M_{NACO}$ as

\small

\begin{eqnarray}\label{nummattot}
M_{NACO} & = &
\left(
\begin{array}{cc}
0.944 & 0.009c_{p}-0.006 \\
-0.005 + 0.008c_{p} + 0.003s_{p} & 0.843-0.015s_{p} \\
-0.003 + 0.004c_{p}-0.007s_{p} & 0.422 + 0.029s_{p} \\
-0.004s_{p}   & -0.057s_{p}
\end{array}
...
\right.
\nonumber
\\
& ...&
\left.
\begin{array}{cc}
0.009s_{p}-0.001 & 0.000 \\
0.366 + 0.015c_{p}& -0.211-0.059s_{p}+0.025c_{p} \\
-0.731-0.029c_{p} & 0.421-0.029s_{p}-0.051c_{p} \\
-0.472 + 0.057c_{p} & -0.816-0.033c_{p}
\end{array}
\right) \;,
\end{eqnarray}

\normalsize

\begin{figure}
   \centering
   \includegraphics[angle=-90,width=8cm]{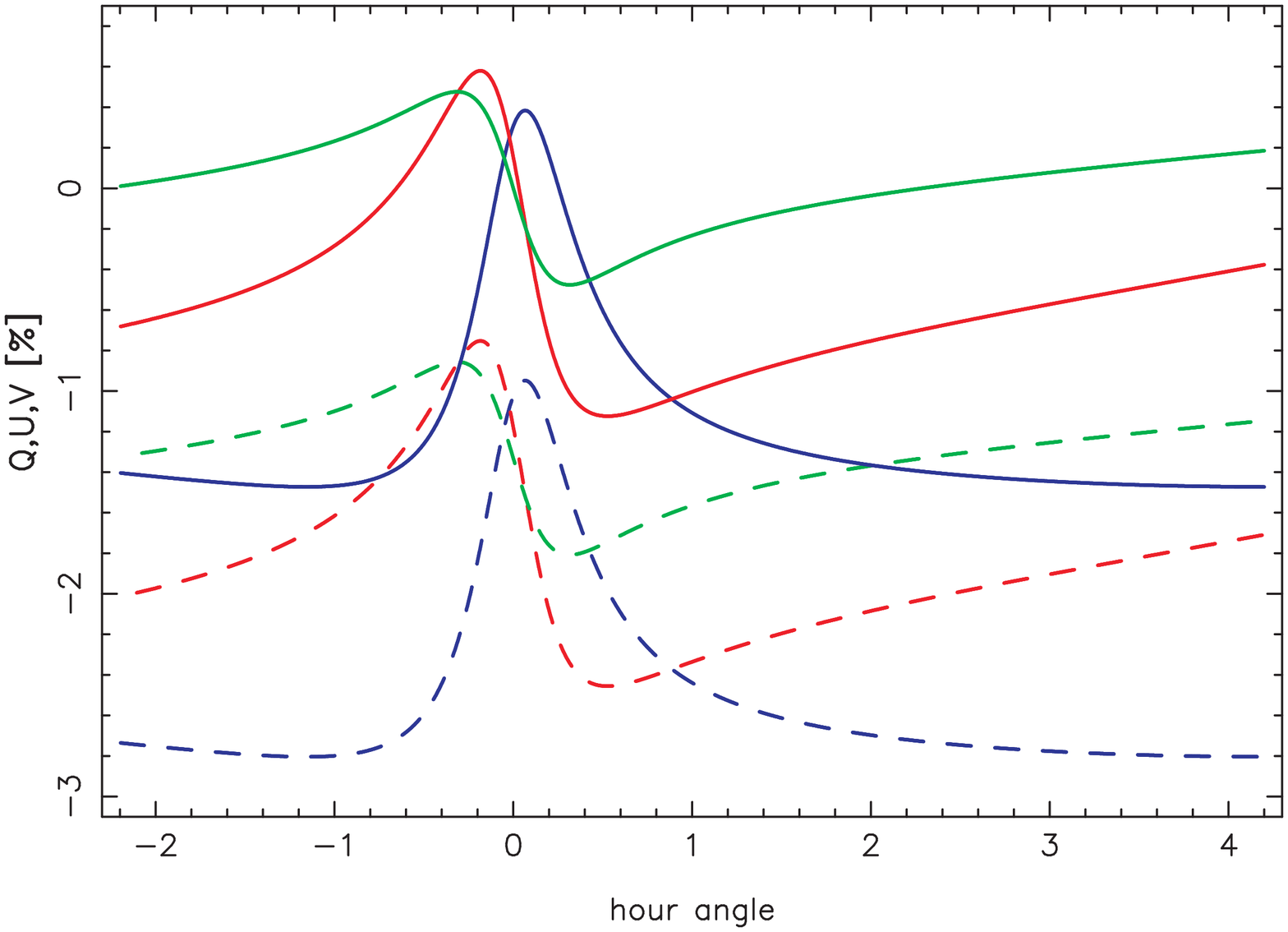}
   \includegraphics[angle=-90,width=8cm]{IPdegha.eps}
   \includegraphics[angle=-90,width=8cm]{IPangha.eps}
      \caption{Upper panel: Instrumental polarization predicted by the model for an unpolarized source corresponding to matrix~(\ref{nummattot}) (solid line) and including the systematic effects of the the analyzer (see section~\ref{polanaly} and Fig.~\ref{flathis} therein, dashed line). The curves show the Stokes parameters $Q$ (blue), $U$ (red), and $V$ (green) as a function of hour angle. Here we considered the most common instrumental setup with Wollaston, HWP, and S13 optics. Middle: Linear (green) and total (red) polarization degree corresponding to the upper Stokes values. Lower panel: Polarization angle.}
         \label{IP}
\end{figure}
\noindent
with $c_{p}=\cos(p)$ and $s_{p}=\sin(p)$ describing the dependency on the parallactic angle. $Tr$ was not included. This last expression gives all necessary information on the cross talks and their dependence on the parallactic angle. The cross talks from $I$ to the linear polarization is on the order of $0.5\%-1\%$ of the total intensity. There are also strong interactions between $Q$ and $U$ (on the order of $40\%$ of the corresponding value $Q$ or $U$ respectively), between $U$ and $V$ (also on the order of $40\%$) and from $V$ to $Q$ (on the order of $20\%$). For not very strongly polarized sources the position-dependent variability of the linear polarization is dominated by the $I \leftrightarrow Q/U$ cross talks.

As an example for the telescope position-dependent behavior of the IP we plot in Fig.~\ref{IP} $Q,U,V$, the linear and total polarization degree, and the polarization angle as functions of hour angle for an unpolarized source at the position of the GC (May, alt = $85.6^{\circ}$ for meridian transit; solid lines in the plots). The IP of NACO reaches about $1.6 \%$ at maximum. Around hour angle zero the IP changes most rapidly, as is expected. Here the polarization degree reaches its minimum and the polarization angle swings to its other extreme. The curves are asymmetrical around zero because of the HWP offset.

\subsection{The polarimetric analyzer} \label{polanaly}

Up to this point we discussed the influence of the optical elements that are located in front of the analyzer. In this section we consider the analyzer itself and its systematic effects on polarimetric measurements.

The main element of the analyzer is a polarizer. After the polarizer the only quantity of interest is intensity, because (for a polarizer with an efficiency sufficiently close to $100 \%$) the polarization is known to be $100 \%$ in the direction of the polarizer (see Eq.~\ref{Qdef}). Thus, the polarizer and the optical elements within the analyzer after the polarizer have to be investigated with respect to their relative attenuation of the different channels.

\subsubsection*{CONICA}

After the Wollaston the optics of CONICA include two more folding mirrors. These mirrors have gold coatings and show different inclinations for S13 and S27. Because after the Wollaston only the intensity of the two channels is important, it is not useful to work with matrix~(\ref{R}), which would describe the change in polarization inside the analyzer. We are interested instead in how the measurement of the Stokes vector of the light in front of the analyzer is affected by the different attenuation of the intensities of both channels inside the analyzer.

For CONICA the relative attenuation factors for the orthogonal channels of the Wollaston are the same for the measurement of $Q$ and $U$, because the arrangement of the optical elements after the Wollaston does not depend on the position of the HWP. It is possible to find a Mueller matrix for this influence.

The matrix to correct for a transmittance difference of the two orthogonal polarimetric channels after the Wollaston can be deduced in the following way:
Let be $I_{1}$ and $I_{2}$ the intensity of the ordinary and the extraordinary beam, respectively. The folding mirrors after the Wollaston attenuate these intensities:
\begin{eqnarray}
I'_{1} & = & r_{1}I_{1} \nonumber \\
I'_{2} & = & r_{2}I_{2} \;,
\end{eqnarray}
with  $r_{1} < 1$ and $r_{2} < 1$ the attenuation factors of the ordinary and the extraordinary beam, respectively. Thus, the measured not normalized Q (other stokes parameters $U$ and $V$ analog) will be
\begin{eqnarray}
Q' & = &  I'_{1}-I'_{2}\nonumber \\
I' & = & I'_{1}+I'_{2} \;.
\end{eqnarray}
The correction matrix applied to the measured Stokes vector $S'$ has to give the vector $S$, which results from the intensities $I_{1}$ and $I_{2}$. Because each of the Stokes parameters $Q$, $U$, and $V$ is determined from a pair of orthogonal Wollaston channels, all of them are affected in the same way, and we can expect e.g. $Q$ to depend on $I'$ and $Q'$ only (analog for $U$ and $V$):
\begin{equation}
Q = xI'+yQ' \;.
\end{equation}
This leads to the equation
\begin{equation}
I'_{1}\left(\frac{1}{r_{1}}-x-y\right)-I'_{2}\left(\frac{1}{r_{2}}+x-y\right) = 0 \;,
\end{equation}
which has the solution
\begin{eqnarray}\label{xy}
x & = & \frac{r_{2}-r_{1}}{2r_{1}r_{2}} \nonumber \\
y & = & \frac{r_{2}+r_{1}}{2r_{1}r_{2}} \;.
\end{eqnarray}
In a similar way we find the matrix elements for $I$. The resulting matrix is the inverse matrix of the transmission matrix $Tr$ of Eq.~\ref{totipmat} and can be written as
\begin{equation} \label{trainv}
Tr^{-1} = \frac{1}{2r_{1}r_{2}} \left(
\begin{array}{cccc}
T_{+} & T_{-} & 0 & 0 \\
T_{-} & T_{+} & 0 & 0 \\
T_{-} & 0 & T_{+} & 0 \\
T_{-} & 0 & 0 & T_{+}
\end{array}
\right) \;,
\end{equation}
with $ T_{\pm} = r_{2} \pm r_{1} $, and by inversion we find
\begin{equation} \label{tra}
Tr = 2r_{1}r_{2} \left(
\begin{array}{cccc}
\frac{T_{+}}{T_{+}^{2}-T_{-}^{2}} & -\frac{T_{-}}{T_{+}^{2}-T_{-}^{2}} & 0 & 0 \\
-\frac{T_{-}}{T_{+}^{2}-T_{-}^{2}} & \frac{T_{+}}{T_{+}^{2}-T_{-}^{2}} & 0 & 0 \\
-\frac{T_{-}}{T_{+}^{2}-T_{-}^{2}} & \frac{T_{-}^{2}}{T_{+}\left(T_{+}^{2}-T_{-}^{2}\right)} & \frac{1}{T_{+}} & 0 \\
-\frac{T_{-}}{T_{+}^{2}-T_{-}^{2}} &
\frac{T_{-}^{2}}{T_{+}\left(T_{+}^{2}-T_{-}^{2}\right)} & 0 & \frac{1}{T_{+}}
\end{array}
\right) \;.
\end{equation}

For the Wollaston (polarization of the ordinary beam orthogonal to the plane of incidence at the first CONICA mirror) the attenuation factors $r_{1}$ and $r_{2}$ can be computed\footnote{Under the assumption of 100\% efficiency of the Wollaston as a polarizer and with the polarization of the ordinary beam perpendicular to the plane of incidence (first gold mirror of CONICA).} from the reflection coefficients for gold for the corresponding angles of incidence (for values see Table~\ref{mat})
\begin{eqnarray}\label{rs}
r_{1} & = & \frac{1}{2}\left(r_{\parallel}^{gold, I}r_{\parallel}^{gold, II}+r_{\perp}^{gold, I}r_{\perp}^{gold, II} \right) \nonumber \\
r_{2} & = & r_{\perp}^{gold, I}r_{\perp}^{gold, II} \;.
\end{eqnarray}

\begin{figure}
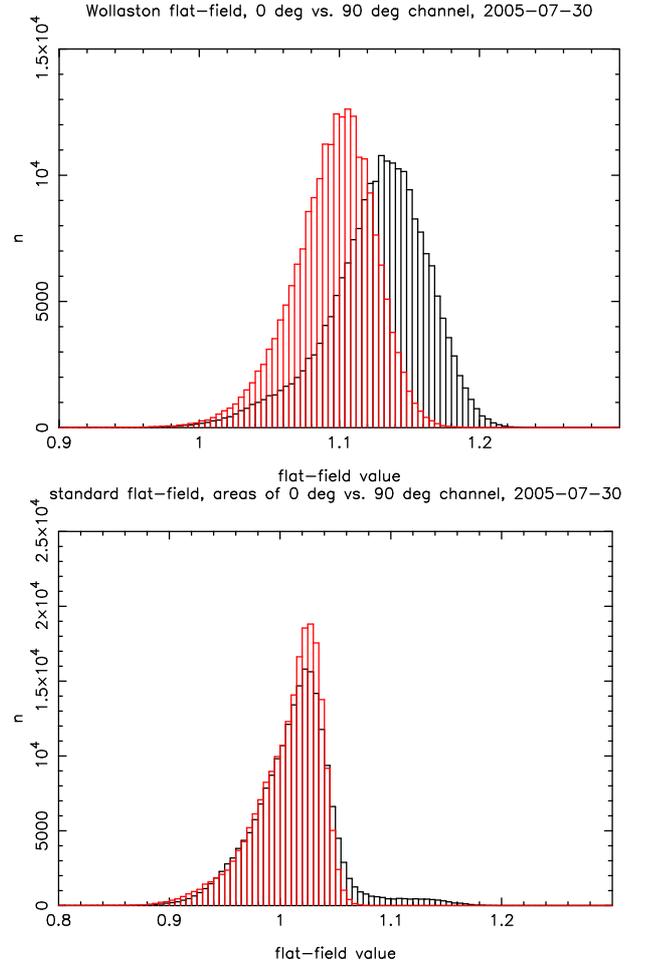

   \centering
   \includegraphics[angle=-90,width=8cm]{flatvar2005a.eps}
   \includegraphics[angle=-90,width=8cm]{flatvar2005b.eps}
      \caption{Histogram of flat-field pixel values in the regions of the orthogonal channels for the S13 polarimetric mask. The upper plot shows the histogram for a detector illuminated by the calibration lamp through the polarizer, the lower one for a simple imaging twilight flat.}
         \label{flathis}
\end{figure}

Unfortunately the information on the attenuation caused by the two mirrors is not sufficient for a correction of the IP of CONICA. The characteristics of the Wollaston, in particular a possibly different transmission of the two beams, are not taken into account. These specifications are not accessible to us, but information on the Wollaston and its transmission is carried by the flat-field that is taken routinely with mask and polarizer. To correct for these transmission differences the relative weighting of the orthogonal channels in the flat-field has to be conserved while normalizing the flat-field. This relative weighting also contains information on the attenuation of the CONICA mirrors after the Wollaston and additionally on the polarization of the calibration lamp that should be zero, but in reality does contribute. The calibration lamp consists of a halogen bulb within a slot of an Ulbricht sphere, which effectively depolarizes the light ($\sim 10^{-3} \% $ remaining). This light is then coupled in by a $45^{\circ}$-tilted gold mirror. This results in a maximum linear polarization of $\sim 1 \%$ of the calibration light.

A comparison between a typical imaging and a Wollaston flat-field is shown in Fig.~\ref{flathis}. Obviously the flat-field with the polarizer shows a significantly different distribution for the detector areas of the two channels, while a flat-field without polarizer exhibits a comparatively homogeneous response in these areas. The ratio of the mean values of the areas that correspond to the extraordinary and the ordinary beam respectively is 1.027 for the flat-field of 2005 (shown in Fig.~\ref{flathis}), whereas the ratio of $r_{2}$ and $r_{1}$ obtained with Eq.~\ref{rs} is only 1.006. Thus, we recommend the correction by the Wollaston flat-field because the influence of the transmission of the Wollaston (which is also different within the field of view of the individual channels) to the polarization can be on the order of 2\% in polarization degree. The dashed lines in Fig.~\ref{IP} show the same model as given by the solid lines, but now including the effects of the analyzer (simulated using Eq.~\ref{totipmat} including $Tr$ with $r_{2}/r_{1}=1.027$). The general behavior of the curves is the same, but the maximum of the IP is now as high as $4\%$ and the angle flips over by full $180^{\circ}$.

We point out here that the polarization of the flat-field calibration light is on the same order as the remaining systematic error of the model described here and probably is the major contribution to the deviation from the standard stars (see section~\ref{gaug}).

\begin{figure}
   \centering
   \includegraphics[width=8cm]{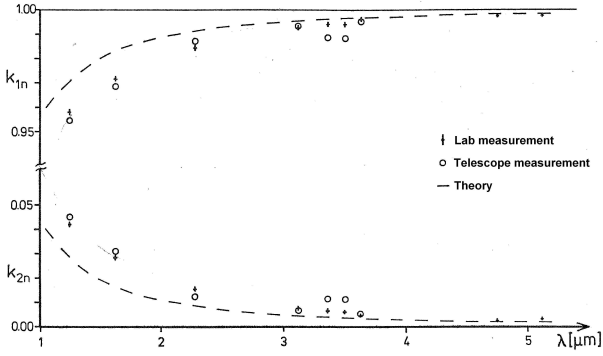}
      \caption{Normalized transmission for polarized light parallel ($k_{2n}$) and orthogonal ($k_{1n}$) to the wires of the grid (Fig. from \citealt{diplhodapp}). Grid period 0.25 $\mu$m, equally spaced on a $\rm{CaF}_{2}$ substrate.}
         \label{effi}
\end{figure}

By using matrix~(\ref{tra}) we assume that the attenuation factors are the same for the measurement of $Q$ as for the measurement of $U$, which is the case for the Wollaston, but not for the wire grid mode. The transmission correction for wire grids is even more complex. For NACO four wire grids were available for observations until and including 2007. The position angles of these grids can be found under the header keyword "INS OPTI4 NAME", which indicates the direction of the wires, thus for the polarization direction that is transmitted, $90^{\circ}$ have to be added.
For this mode one has to consider the more general case of a polarizer with an efficiency $\eta<1$. Fig.~\ref{effi} shows the efficiency as a function of the wavelength. Let $\eta_{1}$ be the efficiency of the polarization direction perpendicular to the wires and $\eta_{2}$ the efficiency parallel to the wires. Then the expressions for $x$ and $y$ in (\ref{xy}) are still valid if we replace $r_{1}$ and $r_{2}$ with $r'_{1}$ and $r'_{2}$, where
\begin{eqnarray}
r'_{1} & = & \eta_{1}r_{1}-\eta_{2}r_{2} \nonumber \\
r'_{2} & = & \eta_{1}r_{2}-\eta_{2}r_{1} \;.
\end{eqnarray}
Here the attenuation factors are not the same for the measurements of the Stokes parameters $Q$ and $U$: $r_{1}$ and $r_{2}$ have to be calculated separately for the two position angle pairs ($0^{\circ}/90^{\circ}$ and $45^{\circ}/135^{\circ}$) of the wire grids according to their the orientation with respect to the mirrors within CONICA. Because in this mode the flat-fields have also to be taken separately, they cannot be used easily to infer information on the relative transmission. The "boot strapping" method described in section~\ref{boot} circumvents these problems.

\subsection{A correction algorithm}

To obtain the true polarization of a source we can apply the inverse of the Mueller matrix to the measured Stokes vector:
\begin{equation}
S = C_{NACO}\times S' = M_{NACO}^{-1}\times S' \;.
\end{equation}
This is possible because every matrix in Eq.~\ref{totipmat} is invertable and therefore $M_{NACO}$ as well. An analytic solution for $C_{NACO}$ can be found in appendix~\ref{appA}.
With NACO we cannot gain information on the circular polarization, and we have to assume  $V'=0$.
However, because in the NIR the circular polarization for most sources can be neglected ($V=0$) and the instrumental circular polarization of NACO can be on the order of $1\%$, the assumption of $V'=0$ introduces an error of this range to the other Stokes parameters when applying the correction matrix $C_{NACO}$. To overcome this problem, an iterative algorithm has to be implemented with the following steps:
\begin{enumerate}
\item Compute the normalized parameters $Q'$ and $U'$ from NACO data; complete the Stokes vector $S'$ with $I'=1$ 
(because the Stokes parameters are normalized by intensity) and set $V'=0$.
\item Apply $C_{NACO}$ (as shown in appendix~\ref{appA}) and obtain a first guess on the 
corrected Stokes vector: $S_{i=0}=C_{NACO}\times S'$.
\item To initiate the $i$-th step define $\hat{S} = (\hat{I},\hat{Q},\hat{U},\hat{V})$  by setting
 $\hat{I}=I_{i-1}$,
 $\hat{Q}=Q_{i-1}$,
 $\hat{U}=U_{i-1}$,
 $\hat{V}=0$.
\item Compute the numerical inverse of $C_{NACO}$ ($C^{-1}_{NACO} = M_{NACO}$) and apply it to $\hat{S}$:
\hspace{0.5cm} $\hat{S'}=C^{-1}_{NACO}\times \hat{S}$.
\item Replace $V'=0$ in $S'$ (defined in step 1) with $\hat{V'}\neq0$ (computed in step 4) and get $S_{i}=C_{NACO}\times S'$.
\item For the next iteration step start from 3.
\end{enumerate}
This algorithm quickly converges to a stable set of Stokes parameters. After 10 iterations the differences in the 
obtained solutions approach the computational uncertainties. It replaces $V'=0$ by a value that guarantees $V$ to be zero.

The algorithm has to be applied for every single frame and its corresponding parallactic angle and rotator position. Often it is necessary to work with mosaics that are obtained by averaging over a number of frames.
Since the differences of the elements of the correction matrices for individual frames are often small within a dataset, it can be a suitable approximation to use the average of the matrices to correct polarimetric data that are obtained from mosaics.

\section{Observations \label{comp}}

In this section we justify the presented model. The comparison with calibration star data shows that we can indeed describe the parallactic angle-dependent IP as it is observed. It also reveals to which accuracy we are able to compensate the systematic effects with our model.

In principle it would be possible to determine the instrumental polarization of an optical train by calibration measurements only: sufficiently bright sources with zero polarization, with linear polarization at $0^{\circ}$ ($U=0$), with linear polarization of an angle close to $45^{\circ}$ ($U\neq0$), and with significant circular polarization (or simply four linearly independent Stokes space elements) would provide the necessary information to solve the equations for the matrix elements. In practice it is not feasible to measure four sources for every parallactic angle, rotator position, and both optics, and in particular it is not possible to measure the Stokes parameter $V$. However, the results of the presented model can be compared with the small number of available and suited standard observations in the ESO archive. In this way it is much easier to obtain a calibration of the IP.

\subsection{The data}

In order to gauge the model for the instrumental polarization we have to test the predictions of the model against standard star observations. In the data archive of ESO plenty of standard star observations are available, mainly of unpolarized standards. Most of these data are difficult to handle because of bad weather conditions, insufficient brightness, or the small number of frames that have been observed. Furthermore polarimetric measurements at different angles have been obtained for these standards by turning the whole instrument instead of the HWP, which changed the IP. Thus, in these cases $Q$ and $U$ have been measured with different instrumental setups. However, three unpolarized stars turned out to be suited to be compared with the prediction of our model. A fourth, the standard RCra88, was even observed with the same observing strategy as Sgr~A*. This star has been studied by \cite{1992ApJ...386..562W}, and it shows a K-band polarization of $(1.8\pm0.1)\%$ at $(95\pm1)^{\circ}$.

The available polarimetric standard star data in the archive do not include longer light curves, which would allow us to investigate the position dependency of the instrumental polarization. For this purpose we use bright stars in the IRS16 cluster of the Galactic center that were observed with a time sampling of about 4 min for Stokes parameters I, Q, and U. 

In order to finally test the systematic effects of different calibration methods on the polarimetric light curves of Sgr~A*, we investigated three of the brightest polarized flares in the framework of the new method. All data sets used for this paper were taken in the Ks-band and are listed in Table~\ref{obs}.

For the GC observations the infrared wavefront sensor of NAOS was used to lock the AO loop on the NIR bright (K-band magnitude $\sim6.5$) supergiant IRS 7, located about 5.6" north of Sgr~A*. For the standard RCra88 the AO was locked on the target itself. For all other standards the AO loop was open.

During the observations the atmospheric conditions (and consequently the AO correction when the loop was closed) were stable enough for high angular resolution photometry and polarimetry (typical coherence time $>2\rm{ms}$). The exposures were jittered by a few arcseconds. All frames were sky-subtracted, flat-fielded, corrected for bad pixels, and aligned with sub-pixel accuracy by a cross-correlation method (\citealt{1999ASPC..172..333D}). For the GC observations point spread functions (PSFs) were extracted from the individual frames with StarFinder (\citealt{2000SPIE.4007..879D}), the images were de-convolved with the Lucy-Richardson (LR) algorithm (which is necessary to counter source confusion and crowding), and a beam restoration was carried out with a Gaussian beam of a FWHM corresponding to the resolution at 2.2$\mu m$ ($\sim60$ mas).

As a preparation for the differential measurements, all channels were aligned to each other on subpixel-scale. Flux densities were measured by aperture photometry. The radius of the apertures for the de-convolved GC data was about 40 mas (3 pixels, S13), about 270 mas (20 pixels, S13) for the isolated RCra88 and about 600 mas (22 pixels, S27) for all other standards. Because of the channel alignment the positions of the apertures were the same for all channels. As a correction for background flux we subtracted the flux measured in apertures where no apparent source is located for each channel. Total intensity for the light curves was obtained by adding the flux of the orthogonal channel and applying a flux density calibration as described in \cite{2010A&A...510A...3Z}. The polarimetric parameters $Q$ and $U$ were then obtained according to Eq.~\ref{Qdef}.

\subsection{Gauging the model: standards and IRS16 stars \label{gaug}}

The GC observations and the measurement of RCra88 used the Wollaston-HWP setup,
whereas for all standard-star observations NACO itself was rotated for the different angles.
In the former case it was possible to compare the data (after correcting for the IP of CONICA by the 
flat-field as described in \ref{totip} and for the offset of the HWP) in $Q$ and $U$ with the predictions of the model.
In the latter case the IP was different for the different angles of the instrument rotation. Thus, a measurement of $Q$ and $U$ under the same circumstances is not available. In these cases we computed the expected difference between the fluxes of both channels for each frame and its instrumental setup and compared it with the data (i.e. we considered each frame as a measurement of $Q$ with a different instrumental setup). For all polarimetric standards the model agrees with the observations with an accuracy of below 0.5 \% (in $Q$ and $Q \& U$ respectively). The standard stars, their reference polarizations (degree and angle), and the systematic deviation of the model from the differential fluxes in the observations are given in Table~\ref{pol}.
The uncertainty of this systematic error was determined taking the median deviation of the data, i.e. the statistical error of the measurements.

 \begin{table}
      \caption[]{Standard stars.}
         \label{pol}
     $$
         \begin{array}{llll}
            \hline
            \hline
            \noalign{\smallskip}
            \rm{Source}      &  P & \phi  & |\rm{model} - \rm{data}|\\
            & [\%] & [\rm{deg}] & [\%] \\
            \noalign{\smallskip}
            \hline
            \noalign{\smallskip}
            \rm{RCra88} & 1.8  & 95 &  (0.2 \pm 0.2)  \;(Q \& U)\\
            \rm{WD 1344}     & 0   &   -   &   (0.3 \pm 0.2)   \;(Q)\\
            \rm{WD 2039-202}     & 0   &   -  &   (0.3 \pm 0.2)   \;(Q)\\
            \rm{HD 109055}     & 0   &   -   & (0.3 \pm 0.2)   \;(Q)\\
            \noalign{\smallskip}
            \hline
          \end{array}
     $$
\tablefoot{Reference polarizations of the standard stars and systematic deviation of the data from the model. For RCra88 the data allowed us to compare both parameters $Q$ and $U$. For all other standards only the differential flux of the orthogonal channels (equivalent to $Q$) for different instrumental setups could be tested.}
   \end{table}

The description of the position-dependent part of the IP can be tested with the long light curves of bright IRS16 sources from 2009. Here a comparison revealed that $k^{alu}$ and $\delta^{sil}$ had to be slightly adjusted to match the shape of the observed light curves in $Q$ and $U$ as described in the captions of Fig.~\ref{mat}. This excursion from the default material constants is within the typical tolerances and can probably be explained with an aging of the aluminum coating.
The time-dependency of the model and the data for Stokes $Q$ and $U$ excellently agree with each other.
Since the apparent\footnote{By "apparent" polarization here and below we mean the polarization that an ideal instrument would measure. The polarization of the stars discussed here is dominated by the interstellar medium.} polarization of sources at the GC is not known with the accuracy of standards, the apparent polarization parameters of these sources are considered to be the free parameters of the fits.

\begin{figure}
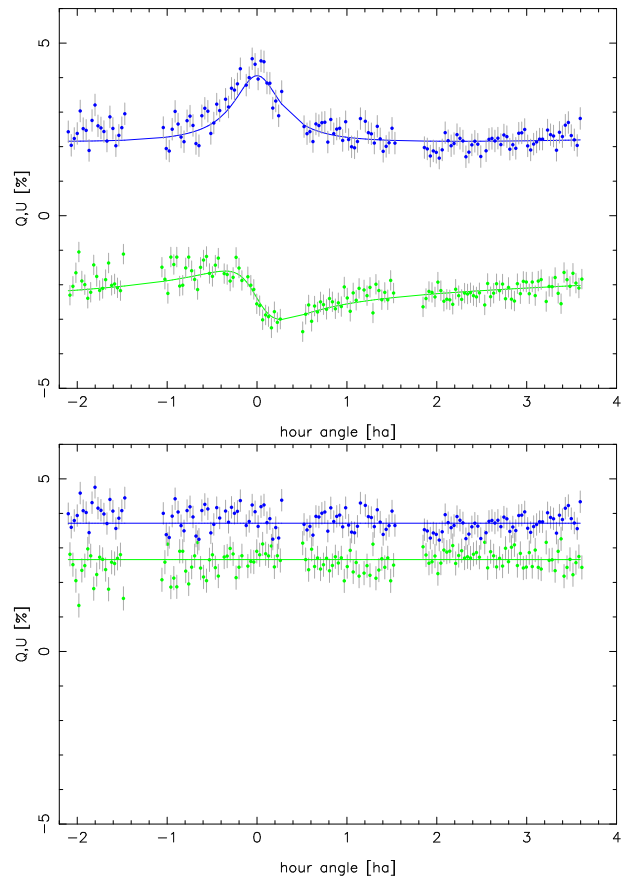

   \centering
   \includegraphics[angle=-90,width=8cm]{quha.eps}
   \includegraphics[angle=-90,width=8cm]{quhacorr.eps}
      \caption{$Q$ (blue) and $U$ (green) parameters as a function of time for IRS16C. 
 The lower panel shows the data corrected for the full instrumental polarization.}
         \label{irsplots}
\end{figure}

Fig.~\ref{irsplots} shows model and data for IRS16C (In Figures~\ref{irsplots2} we show the same model vs. data comparison for IRS16NW, IRS16CC, and S67).
In this figure $Q$ (blue) and $U$ (green) parameters are shown as a function of hour angle.
The solid line is the best $\chi^2$-fit of the model to the 2009 data that is already corrected for the HWP offset (upper panel). The lower panel shows the data corrected for
the full instrumental polarization.
The solid line here describes $Q$ and $U$ corresponding to the apparent polarization of the source. The big difference between the measured and the corrected $U$ parameter results mainly from the opposite sense of rotation of the HWP with respect to the sky. The larger deviation of the model from the data points at the beginning of the night are due to the weather conditions and the AO performance, which were not as stable as later. The fitting of the model is weighted toward the end of the night.

For all four fitted sources the errors of the single data points are about $0.3 \%$ for $Q$ and $U$ and $\chi^2/\rm{dof} = 1.1$,
where $p$ and $\phi$ (apparent polarization), and $k^{alu}$ and $\delta^{sil}$ 
(material constants, see Table~\ref{mat}) are considered to be the free parameters.
The apparent polarizations as best $\chi^{2}$ fitting results are listed in Table~\ref{polvar} for all four IRS16 sources obtained from 2009 data.
In this table the error of the polarization degree is about $0.8 \%$, the error of the angle about $3^{\circ}$.
For comparison we list the results of \cite{1999ApJ...523..248O}.
Here $p$ has in average an error of $2 \%$, and the angle uncertainty is about $18^{\circ}$.
Additionally we give the average polarization of the central arcsecond around Sgr~A* for stars with $m_{k} \ge 13$. Stars of this brightness in the near surrounding of Sgr~A* have commonly be used as calibration stars for the "boot strapping" calibration described in section~\ref{boot}. We emphasize that to our knowledge the results presented here are the first polarimetric measurements of sources within the central pc of the GC since \cite{1977ApJ...216..271K} that are independently calibrated with a method that goes beyond "boot strapping" procedures (see section~\ref{boot}). The polarization of the sources in Table~\ref{polvar} compares well with the polarization found by \cite{1977ApJ...216..271K} for the central region of the GC and can therefore be explained by the galactic foreground polarization as discussed in \cite{1977ApJ...216..271K} and \cite{1999ApJ...523..248O}.

The presented model enables us to correct the IP with an accuracy better than $1 \%$ in polarization degree and better than $5^{\circ}$ in polarization angle for polarization degrees $\ge 4\%$. These errors are deduced from the light curves of the IRS16 sources that exhibit comparably small statistical errors for $Q$ and $U$, and the systematical deviation of $Q$ and $U$ of $0.4 \%$ for the standard stars (see Table~\ref{pol}). The time variability (i.e. the relative behavior) of the IP in polarization degree can be described with an accuracy of a few tenths of a percent. Thus, after correction of the IP, remaining variability with amplitudes of $1 \%$ or more in linear polarization is caused by intrinsic variability and statistical errors and is not a feature of the IP.

 \begin{table}
      \caption[]{Apparent polarization of sources at the GC.}
         \label{polvar}
     $$
         \begin{array}{llllll}
            \hline
            \hline
            \noalign{\smallskip}
\rm{Source}   & \multicolumn{2}{l}{\textnormal{This paper}}  & \multicolumn{2}{l}{\textnormal{\cite{1999ApJ...523..248O}}} & m_{k}\\
	    \noalign{\smallskip}
               &  P & \phi & P & \phi &  \\
               &  [\%] & [\rm{deg}] & [\%] & [\rm{deg}]  & \\
            \noalign{\smallskip}
            \hline
            \noalign{\smallskip}
            \rm{IRS\;16C\;(S96)}      &      4.6  & 17.8   & 4.0  & 35  & 9.55   \\
            \rm{IRS 16CC}      &       5.4  & 15.5  & 6.1  & 54   & 10.15 \\
            \rm{IRS\;16NW\;(S95)}      &       5.9  & 12.0  & 4.6  & 24   & 9.86 \\
            \rm{S67}      &       5.2  & 17.8 & - & - & 12.10 \\
            \noalign{\smallskip}
            \hline
            \noalign{\smallskip}
            \rm{central \hspace{0.1cm} arcsec}      &       5.3  & 27   & \\
            \noalign{\smallskip}
            \hline
         \end{array}
     $$
\tablefoot{K-band magnitudes are taken from \cite{1999ApJ...523..248O} and \cite{2009ApJ...692.1075G}.}
   \end{table}

\section{Comparison of different calibration methods \label{cacom}}

In this section we describe two other common calibration methods. We investigate their ability to correct the IP of NACO. Both calibration methods have shortcomings, and we give a quantitative analysis of the remaining systematic errors. We investigate the influence of the "boot strapping" method that was commonly used for calibrating light curves of Sgr~A* on the evaluation of time series of the polarimetric parameters. But first we discuss the "channel switch" method, which enables us to propose an observing strategy that aims at high-accuracy polarimetry, but not at high-time resolution.

\subsection{The "channel switch" method}
\label{channelswitch}

A common method to treat instrumental polarization is to observe orthogonal channels at $0^{\circ}/90^{\circ},45^{\circ}/135^{\circ},90^{\circ}/0^{\circ},135^{\circ}/45^{\circ}$. We assume that the IP affects each of the orthogonal channels factorially, and that the particular factor does not change if the instrument is turned by $90^{\circ}$:
\begin{equation}\label{eqfac}
Q_{0^{\circ}} = \lambda I(0^{\circ})-\sigma I(90^{\circ}) \;\; \rm{and} \;\; Q_{90^{\circ}} = \lambda I(90^{\circ})-\sigma I(0^{\circ}) \;,
\end{equation}
with $\lambda$ and $\sigma$ the factors introduced by the IP (for $U$ analog equations). The true Stokes parameters then can be obtained from the difference of the Stokes parameters derived from the corresponding images $0^{\circ}/90^{\circ}$ and $90^{\circ}/0^{\circ}$, and $45^{\circ}/135^{\circ}$ and $135^{\circ}/45^{\circ}$ respectively, and with Eq.~\ref{Qdef} we get
\begin{equation}\label{csc}
Q_{corr} = \frac{Q_{0^{\circ}}-Q_{90^{\circ}}}{I_{tot,0}+I_{tot,90}} = \frac{(\lambda +\sigma)(I(0^{\circ})-I(90^{\circ}))}{(\lambda +\sigma)(I(0^{\circ})+I(90^{\circ}))} = Q
\end{equation}
(for U analog).

The assumption of equal factors in Eq.~\ref{eqfac} is not appropriate for every case. Obviously the influence of a transmission matrix like $Tr$ of Eq.~\ref{tra} can be fully corrected by "channel switching", whereas for the matrices~(\ref{R}) and~(\ref{stokesrot}) systematical errors remain. To quantify the remaining effects for a given Mueller matrix $M$ we calculate the Stokes vectors
\begin{equation}
S'_{0^{\circ}}= M \times S \;\; and \;\; S'_{90^{\circ}}= M \times T(90^{\circ}) \times S
\end{equation}
with $S$ the Stokes vector of the incoming light and $S'_{0^{\circ}}$ and $S'_{90^{\circ}}$ the measured Stokes vectors of the image pairs $0^{\circ}/90^{\circ},45^{\circ}/135^{\circ}$ and $90^{\circ}/0^{\circ},135^{\circ}/45^{\circ}$ respectively. For matrix~(\ref{R}) we find
\begin{equation}
Q_{corr} = Q \;\; \rm{and} \;\; U_{corr} = \frac{\sqrt{r_{\perp}r_{\parallel}}\cos(\delta)}{2(r_{\perp} + r_{\parallel})}U \;,
\end{equation}
with $Q$ and $U$ the proper Stokes parameters of the incoming light. For matrix~(\ref{stokesrot}) we find
\begin{eqnarray}
Q_{corr} & = & \cos(2p)Q+\sin(2p)U \;\; \rm{and} \nonumber \\
U_{corr} & = & \cos(2p)U-\sin(2p)Q \;.
\end{eqnarray}
That is, a rotation of the angle reference as the one caused by the HWP offset cannot be corrected at all, while for a metallic reflection $U$ is still affected by a factor that depends on the material constants.

Furthermore this method can only correct instrumental effects for the part of the instrument that is rotated by $90^{\circ}$. Because in our case the rotated part is the analyzer (a rotation of the HWP is equivalent to a rotation of the analyzer), this procedure only corrects for the IP of the light train after the HWP. Another possibility is to rotate NACO with the rotation adapter as a whole, as was done for a number of observations in the ESO archive. In both cases the IP of M3 remains uncorrected. Nevertheless, using the "channel switch" method, the effects of the analyzer and the flat-fielding, which are difficult to quantify as described in section~\ref{polanaly}, and even parts of the IP of NAOS can be eliminated. We propose the following strategy.

To archive the best attainable accuracy we propose to realize the $45^{\circ}$-switching between $Q$ and $U$ with the HWP, and the $90^{\circ}$-switching with a rotation of the entire instrument (NACO). We now consider only the effects after M3, i.e. the system $Tr \; \times \; T(-\beta) \times T(90^{\circ}) \times M_{NAOS} \times T(-90^{\circ})$. Then the numerical analysis shows that Eq.~\ref{csc} gives
\begin{eqnarray}
Q_{corr} & = & 0.894Q -0.387U \nonumber \\
U_{corr} & = & 0.447Q +0.775U \;.
\end{eqnarray}
The situation gets much easier if we measure at offset corrected angles (without HWP offset, the angles can be obtained from Table~\ref{enc}). Then we get
\begin{eqnarray}
Q_{corr} & = & Q \nonumber \\
U_{corr} & = & 0.865U \;.
\end{eqnarray}
Here we just have to additionally correct for the factor in $U$ and the IP of M3, which reduces the number of free parameters of the model to 3. The effects of all the parts after M3 can be reduced like this to one factor.

This method in particular eliminates the mentioned uncertainties of the parameters of the analyzer. Since the description of the parallactic angle-dependent part caused by M3 proved to be very accurate (a few tenth of a percent in polarization degree) in comparison to the remaining systematic deviations between model and standards (about $1\%$ in polarization degree), we expect that the proposed method will allow us to improve the accuracy by about a factor ten. We will investigate this in future calibration runs. We emphasize here that the "channel switch" method needs more than twice the time to obtain one set of Stokes parameters and is not suited for Sgr~A* and its fast variability with time scales down to a few minutes. Another disadvantage is the restriction of the field-of-view, which is caused by the rotation of NACO with the rotation adapter.

\subsection{The "boot strapping" method}
\label{boot}
For time-resolved measurements in the crowded Galactic center field we have commonly used 
a "boot strapping" method to calibrate the polarization data
(e.g. \citealt{2006A&A...450..535E,2006A&A...455....1E}, \citealt{2010A&A...510A...3Z}).  This method has also been successfully applied to wire-grid data (e.g. \citealt{1995ApJ...445L..23E}, \citealt{1999ApJ...523..248O}). In the presence of crowded fields with many weakly and only a few strongly polarized sources, it has the advantage of being applicable without the availability of extensive data on calibrator sources. Below we describe this method and investigate its uncertainties in detail.

For the polarimetric "boot strapping" calibration of the light curves of Sgr~A*, each channel is flux-density-calibrated with reference stars in a region of 2 arcseconds diameter surrounding Sgr~A* assuming total intensity brightness for each star. The sums of the orthogonal channels for $0^{\circ}$ and $45^{\circ}$ are averaged and taken as total intensity. With this total intensity and the galactic foreground polarization of $4\% @ 25^{\circ}$ (\citealt{1977ApJ...216..271K}) one obtains with Eq.~\ref{chan} the expected flux densities for each star and channel. These flux densities are then compared with the time-averaged fluxes of the light curves of each star and channel, and a correction factor for each channel is obtained by averaging over all stars.
Following this procedure the stars in the near surrounding of the GC show in average the foreground polarization and every source with similar polarization is calibrated. The value of $4\% @ 25^{\circ}$ is an average for the sources toward the central arcsecond that has been measured by \cite{1977ApJ...216..271K} with arcsecond resolution. We could confirm this measurement by our independent calibration, which results in an average of $5.3\% @ 27^{\circ}$ for the central arcsecond (Table~\ref{polvar}), which is equivalent within the errors.

By using this calibration procedure one assumes that the IP affects the measurement by introducing different weighting factors to the flux measurements of the four polarimetric channels, very similar to the assumption in Eq.~\ref{eqfac}. Here these factors are considered to be independent of the polarization of the considered source (whereas the factors for the "channel switching" are assumed to be independent of an instrument rotation). In particular it cannot correct for an angular offset like the one caused by the HWP if the calibrator's polarization angle is significantly different from the polarization angle of the source one aims to calibrate (this offset corresponds to a $Q \leftrightarrow U$ cross talk, and the correction factors for each channel in this case depend on the direction of the linear polarization). We intend to answer the question of how this systematic error of the "boot strapping" calibration influences the variability of the polarimetric light curves of Sgr~A*.

\begin{figure}
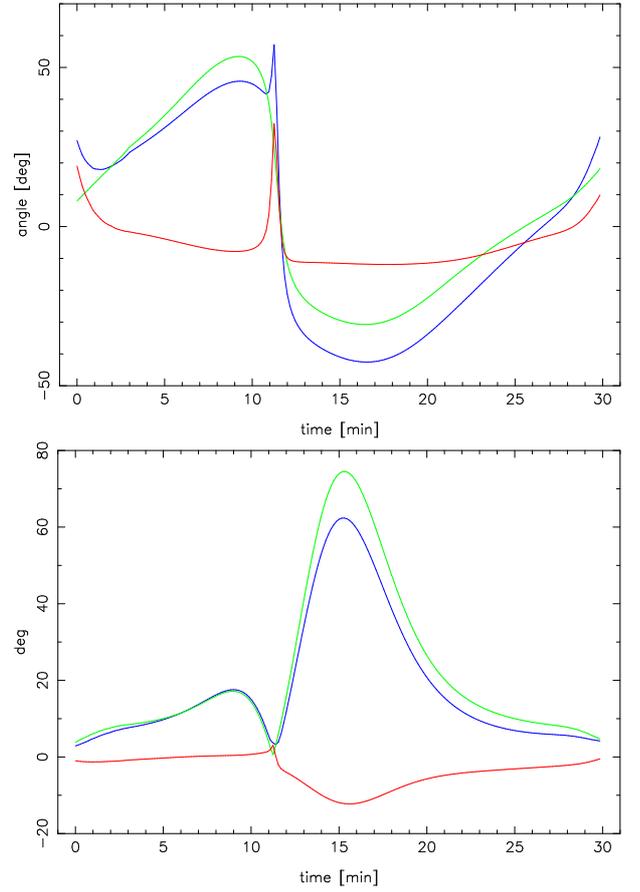

   \centering
   \includegraphics[angle=-90,width=8cm]{poltoterrora.eps}
   \includegraphics[angle=-90,width=8cm]{poltoterrorp.eps}
      \caption{Simulation of systematic calibration artefacts in light curves. The green curves show a typical pattern for polarization degree and angle as a function of time (\citealt{2010A&A...510A...3Z}). The blue curves show the same data calibrated by the described "boot strapping" method. The difference between both is displayed in red.}
         \label{sim}
   \end{figure}

First we investigate the systematic errors for a theoretical light curve pattern as deduced in \cite{2010A&A...510A...3Z} for a polarized orbiting hot spot in an accretion disk around Sgr~A*. In Fig. 37 of \cite{2010A&A...510A...3Z} an apparent view of a hot spot in a Keplerian orbit at the innermost stable circular orbit (ISCO) of a spinning black hole (with spin parameter of 0.5) is shown. The model predicts that the observer witnesses a magnification in flux according to lensing and boosting effects. The polarization angle on the observer's sky sweeps shortly before the total flux reaches its maximum, while the degree of polarization follows this maximum (Fig.~\ref{sim}). The existence this a pattern is an indicator for the strong gravitational regime and can be used as a tool for measuring the spin of a black hole (\citealt{2010A&A...510A...3Z}). While this model predicts the described pattern as a function of the normalized orbital time scale, we here set the orbital time scale to 30 min (as observations of Sgr~A* suggest) and the center of the pattern (here at t=15~min) to hour angle zero, where the variability of the IP is strongest. We compute the Stokes vector for a source that shows the apparent polarization through the  foreground of $5.3\%@27^{\circ}$ (see Table~\ref{polvar}), apply the Mueller matrix for NACO to this vector, and deduce the normalized flux in each channel. These fluxes are then compared with the expected fluxes for a source of $4\% @ 25^{\circ}$ (without IP) as assumed in previous publications (e.g. \citealt{2010A&A...510A...3Z}, \citealt{2006A&A...458L..25M}).
For each time the obtained correction factors are applied to the channel fluxes that have been calculated from the theoretical polarization pattern; this pattern had to be transformed before with $M_{NACO}$ to describe the actual measurement at the detector as predicted by our IP model.

The resulting light curves are shown in Fig.~\ref{sim}. The peak values of the polarization degree are systematically underestimated by about $10\%$ at an expected peak of $70\%$. Typical deviations are on the order of $<$5\% for the degree of polarization and $\le13^{\circ}$ for the polarization angle. However, the angle mainly shows the expected HWP offset of about $13^{\circ}$, while other polarization effects of the instrument have much smaller influences. Indeed, the resulting light curves look very much the same if the simulation only takes the HWP offset into account. For very low polarization degrees the angle is ill defined. At these states real data with white noise contribution do not allow for detecting significant polarized flux, and therefore the interpretation of the polarization angle is not possible in either way. The overall behavior of the variability is conserved. A compensation for the HWP offset during observations would eliminate almost all the effects introduced by this calibration method.

\subsection{Effects on time-resolved polarimetric measurements of Sgr~A*}

As a final step we investigate the difference of the calibration methods with respect to the observed polarimetric data of Sgr~A*. For comparison a proper error estimate of the individual data points is needed. For the common "boot strapping" method the error was deduced from (1) the statistical variations of a comparison star near Sgr~A* for each channel after flux density calibration, and (2) the standard deviation of the correction factors for the calibration stars. This was performed by Gaussian error propagation for $p$ and $\phi$. Since the new calibration does not include a flux density calibration, one has to estimate the statistical error of the photometry from the ADU (analog to digital conversion units) counts of a comparison star. Here it is important to first eliminate the correlated fluctuations of both orthogonal channels. This is achieved by subtracting one channel from the other after scaling the subtracted channel in a way that the averages of both channels are the same\footnote{For comparison stars of low polarization the scaling is not crucial.}. The standard deviation of this difference is a good error estimate for the difference in flux between both channels (and the total flux as well), and can be propagated again.

\begin{figure}
   \centering
   \includegraphics[angle=-90,width=8.0cm]{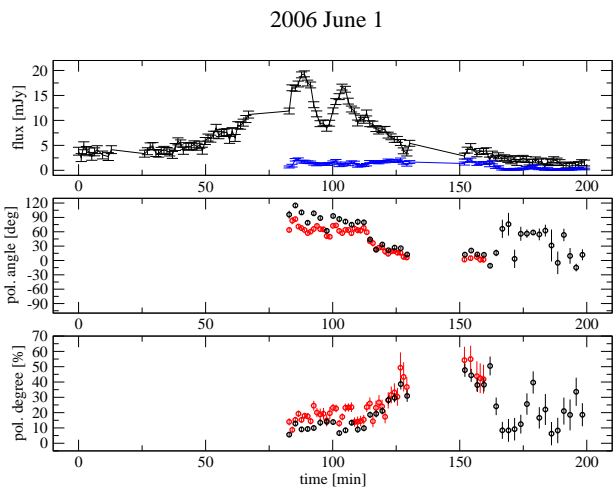}
      \caption{Total intensity (upper panel), polarization angle (middle panel), and polarization degree (lower panel) of Sgr~A* as a function of time (2006 June 1). The black data points are the result of the new calibration method, the red points are obtained by the previously used "boot strapping" method as published in \cite{2010A&A...510A...3Z}. The blue points show polarized flux ($p\times I$). The meridian transit occurred at 32.4 min, 50.4 min before the polarimetric measurements started.}
         \label{lc1}
\end{figure}

As a result we obtain Figures~\ref{lc1}, \ref{lc2}, and \ref{lc3}.
In general both calibration methods show very similar results within the statistical uncertainties of the measurements.
The new calibration shows a trend toward smaller polarization degrees, and the polarization angle shows a small systematic
offset as discussed. Generally, the polarization angle is not well defined for small degrees of polarization.
The comparisons in Figures~\ref{lc1}, \ref{lc2}, and \ref{lc3} show that for total Ks-band intensities above 4~mJy and 
polarized fluxes above 1~mJy the results of both polarization calibration methods are virtually identical. Only states of Sgr~A* that agree with these conditions have been interpreted in the framework of a relativistic modeling (\citealt{2010A&A...510A...3Z}, \citealt{2006A&A...455....1E}, \citealt {2006A&A...458L..25M,2006A&A...460...15M}).

\section{Conclusions \label{concl}}

We presented a detailed analysis of the polarization calibration of the
ESO VLT NAOS/CONICA system in the Ks-band.
Using the Stokes/Mueller formalism for metallic reflections we introduced a polarization model of the 
camera/telescope system that excellently agrees with the measurements obtained on
calibrator sources and sources in the Galactic center.
We can qualitatively and quantitatively reproduce the instrumental polarization and show that a polarization angle offset of $ (13.2 \pm 0.3)^{\circ}$ has to be taken into account for NACO data observed before autumn 2009.
The model presented here enables us to correct for the instrumental polarization of NACO with an accuracy better than $1\%$ in Ks-band. Our approach allows us to extend the description of the IP to other wavelengths easily. It can also be applied to other telescopes and Nasmyth focus instruments. Additional calibration observations may allow for an even better accuracy.
The IP in Ks-band amounts to a maximum of 4\%.
We showed that the "boot strapping" method, which can efficiently be used in the crowded Galactic center field, yields the same results within the statistical uncertainties in bright flare phases when compared to the more exact and elaborate polarization model.

In summary our investigation shows that for sources with statistical errors in $Q$ and $U$ significantly smaller than 0.5 \% the polarization can be measured with an accuracy of better than 1\% in polarization degree. The accuracy of the polarization angle in these cases is $\le 5^{\circ}$ for polarization degrees $\ge 4 \%$. For weaker sources like Sgr~A* the accuracy of the polarization measurements is dominated by the statistical errors.
We find that for total intensity fluxes above 4~mJy and polarized fluxes larger than 1~mJy the previously obtained results on the polarized emission from Sgr~A* are unaffected by our new findings. Finally we proposed a calibration strategy that probably will allow for observations with an accuracy of a few tenths of a percent in the future. For this strategy a compensation for the HWP offset while observing is essential. On the other hand without offset even almost all of the effects introduced by the "boot strapping" calibration method would be eliminated, too.

\begin{acknowledgements}
For their friendly advice and comments we thank A. Stolte, W. Brandner, C. Hummel, A. Witzel, E. Angelakis, N. Marchili, W. Huchtmeier and the ESO user support. Special thanks to Balzers Optics for their congenial help. This paper is based on observations made with the European Southern Observatory telescopes obtained from the ESO/ST-ECF Science Archive Facility. Part of this work was supported by Deutsche Forschungsgemeinschaft, DFG, via grant SFB 494. M. Zamaninasab was member of the International Max Planck Research School (IMPRS) for Astronomy and Astrophysics at the MPIfR Bonn and the Universities of Bonn and Cologne. R. Sch\"odel acknowledges the Ram\'{o}n y Cajal programme by the Spanish Ministry for Science and Innovation. N. Sabha acknowledges support from Bonn-Cologne Graduate School of Physics and Astronomy (BCGS). V. Karas acknowledges the Czech Science Foundation (ref. 205/07/0052). Part of this work was supported by the: COST Action MP0905 Black Holes in a violent Universe.

\noindent
Finally we want to thank the anonymous referee for his helpful comments.
\end{acknowledgements}

\bibliographystyle{aa}
\bibliography{mybib}{}

\begin{appendix}

\section{Correction matrices for the optical components of NACO \label{appA}}

In this appendix we give an analytic solution for the correction matrix $C_{NACO}$ in a way that makes it easy to program it. This solution was obtained by the inverse of the matrices in Eq.~\ref{totipmat}, which we also give here. The inverse of the rotation matrix in Stokes space (see Eq.~\ref{stokesrot}) can be obtained by changing the sign of the parameter $p$. The inverse of the Mueller matrix for metallic reflection~(\ref{R}) is given by
\begin{equation}
R^{-1} = \left(
\begin{array}{cccc}
r_{+} & r_{-} & 0 & 0 \\
r_{-} & r_{+} & 0 & 0 \\
0 & 0 & \Delta_{c} & \Delta_{s} \\
0 & 0 & -\Delta_{s} & \Delta_{C}
\end{array}
\right) \;,
\end{equation}
with variables analog to \ref{rs2}. With the inverted Mueller matrices we are able to give an analytic correction matrix that transforms the measured Stokes vector into the Stokes vector on the sky (with the orientation east of north)
\begin{equation}
S = C_{NACO}\times S' = M_{var}^{-1}\times M_{con}^{-1}\times S' \;,
\end{equation}
with $M_{var}$ the position depending and $M_{con}$ the constant part of the Mueller matrix for NACO
\small
\begin{equation}
M_{var}^{-1} = \left(
\begin{array}{cccc}
r_{+}^{alu} & -ac_{p}-bs_{p} & as_{p}-bc_{p} & 0 \\
r_{-}^{alu}c_{p} & ec_{p}^{2}+fs_{p}c_{p}+t & fc_{p}^{2}-es_{p}c_{p}+u & -\Delta_{s}^{alu}s_{p} \\
r_{-}^{alu}s_{p} & es_{p}c_{p}-fc_{p}^{2}+v & ec_{p}^{2}+fs_{p}c_{p}-w & \Delta_{s}^{alu}c_{p} \\
0 & dc_{p}-gs_{p} & -ds_{p}-gc_{p} & \Delta_{c}^{alu}
\end{array}
\right) \;,
\end{equation}
with

\normalsize
\begin{minipage}[b]{3 cm}
\begin{eqnarray}
s_{p} & = & \sin(2p) \nonumber \\
c_{p} & = & \cos(2p) \nonumber \\
s_{\beta} & = & \sin(2\beta) \nonumber \\
c_{\beta} & = & \cos(2\beta) \nonumber \\
a & = & -r_{-}^{alu}\cos(2\alpha) \nonumber \\
b & = & r_{-}^{alu}\sin(2\alpha) \nonumber \\
d & = & -\Delta_{s}^{alu}\sin(2\alpha) \nonumber \\
g & = & \Delta_{s}^{alu}\cos(2\alpha) \nonumber
\end{eqnarray}
\end{minipage}
\begin{minipage}[b]{5.4 cm}
\begin{eqnarray}
v & = & -r_{+}^{alu}\sin(2\alpha) \nonumber \\
w & = & r_{+}^{alu}\cos(2\alpha) \nonumber \\
t & = & -\Delta_{c}^{alu}\cos(2\alpha) \nonumber \\
u & = & \Delta_{c}^{alu}\sin(2\alpha) \nonumber \\
e & = & \left(r_{+}^{alu}+\Delta_{c}^{alu}\right)\cos(2\alpha) \nonumber \\
f & = & -\left(r_{+}^{alu}+\Delta_{c}^{alu}\right)\sin(2\alpha)
\end{eqnarray}
\end{minipage}

\vspace{0.3 cm}
\noindent and

\small
\begin{equation}
M_{con}^{-1} = \left(
\begin{array}{cccc}
T_{+}k-T_{-}l\left( c_{\beta}+s_{\beta} \right) & T_{-}k-T_{+}lc_{\beta} & -T_{+}ls_{\beta} & 0 \\
-T_{+}l+T_{-}k\left( c_{\beta}+s_{\beta} \right) & -T_{-}l+T_{+}kc_{\beta} & T_{+}ks_{\beta} & 0 \\
T_{-}\left[m\left(c_{\beta}-s_{\beta}\right)-n\right] & -T_{+}ms_{\beta} & T_{+}mc_{\beta} & -T_{+}n \\
T_{-}\left[n\left(c_{\beta}-s_{\beta}\right)+m\right] & -T_{+}ns_{\beta} & T_{+}nc_{\beta} & T_{+}m
\end{array}
\right)
\end{equation}
with

\normalsize
\begin{minipage}[b]{3.8 cm}
\begin{eqnarray}
k & = & \left(r_{+}^{sil}\right)^{2}+\left(r_{-}^{sil}\right)^{2} \nonumber \\
l & = & 2r_{+}^{sil}r_{-}^{sil} \nonumber \\
m & = & \left(\Delta_{c}^{sil}\right)^{2}-\left(\Delta_{s}^{sil}\right)^{2} \nonumber
\end{eqnarray}
\end{minipage}
\begin{minipage}[b]{4.6 cm}
\begin{eqnarray}
n & = & 2\Delta_{c}^{sil}\Delta_{s}^{sil} \nonumber \\
T_{\pm} & = & r_{2} \pm r_{1}
\end{eqnarray}
\end{minipage}

\vspace{0.3 cm}
\noindent and with

\begin{minipage}[b]{3 cm}
\begin{eqnarray}
r_{\pm}^{alu} & = & \frac{1}{2}\frac{r_{\parallel}^{alu} \pm r_{\perp}^{alu}}{r_{\parallel}^{alu}r_{\perp}^{alu}} \nonumber \\
\Delta_{c}^{alu} & = & \frac{\cos(\delta^{alu})}{\sqrt{r_{\parallel}^{alu}r_{\perp}^{alu}}} \nonumber \\
\Delta_{s}^{alu} & = & \frac{\sin(\delta^{alu})}{\sqrt{r_{\parallel}^{alu}r_{\perp}^{alu}}} \nonumber
\end{eqnarray}
\end{minipage}
\begin{minipage}[b]{5.4 cm}
\begin{eqnarray}
r_{\pm}^{sil} & = & \frac{1}{2}\frac{r_{\parallel}^{sil} \pm r_{\perp}^{sil}}{r_{\parallel}^{sil}r_{\perp}^{sil}} \nonumber \\
\Delta_{c}^{sil} & = & \frac{\cos(\delta^{sil})}{\sqrt{r_{\parallel}^{sil}r_{\perp}^{sil}}} \nonumber \\
\Delta_{s}^{sil} & = & \frac{\sin(\delta^{sil})}{\sqrt{r_{\parallel}^{sil}r_{\perp}^{sil}}} \nonumber
\end{eqnarray}
\end{minipage}
\begin{eqnarray}
r_{1} & = & \frac{1}{2}\left(r_{\parallel}^{gold, I}r_{\parallel}^{gold, II}+r_{\perp}^{gold, I}r_{\perp}^{gold, II} \right) \nonumber \\
r_{2} & = & r_{\perp}^{gold, I}r_{\perp}^{gold, II} \label{rs2} \;.
\end{eqnarray}

\vspace{0.3 cm}
\noindent The parameters are

\begin{eqnarray}
p & = & \hspace{0.2 cm} \rm{parallactic} \hspace{0.2 cm} \rm{angle} \nonumber \\
\alpha & = & \hspace{0.2 cm} \rm{rotator} \hspace{0.2 cm} \rm{angle} \nonumber \\
\beta & = & { 13.2^{\circ}} \nonumber \\
r_{\parallel}^{alu} & = & 0.962622 \nonumber \\
r_{\perp}^{alu} & = & 0.981133 \nonumber \\
\delta^{alu} & = & 176.028^{\circ} \nonumber \\
r_{\parallel}^{sil} & = & 0.98272 \nonumber \\
r_{\perp}^{sil} & = & 0.98872 \nonumber \\
\delta^{sil} & = & 165^{\circ} \nonumber \\
r_{\parallel}^{gold, I} & = & 0.972664/0.977011  \hspace{0.2 cm} \rm{(S13/27)} \nonumber \\
r_{\perp}^{gold, I} & = & 0.98484/0.981932  \hspace{0.2 cm} \rm{(S13/27)} \nonumber \\
r_{\parallel}^{gold, II} & = & 0.979588/0.97865  \hspace{0.2 cm} \rm{(S13/27)} \nonumber \\
r_{\perp}^{gold, II} & = & 0.979642/0.980538  \hspace{0.2 cm} \rm{(S13/27)} \;.
\end{eqnarray}

These matrices are not normalized, because it would make them more difficult to read. This just effects total intensity, the normalized Stokes parameters remain unaffected. To switch off matrix $Tr^{-1}$ after a flat-field correction as described in \ref{polanaly} set $r_{1}=r_{2}$.
\newpage
\section{Tables and figures}

\begin{figure}[h!]
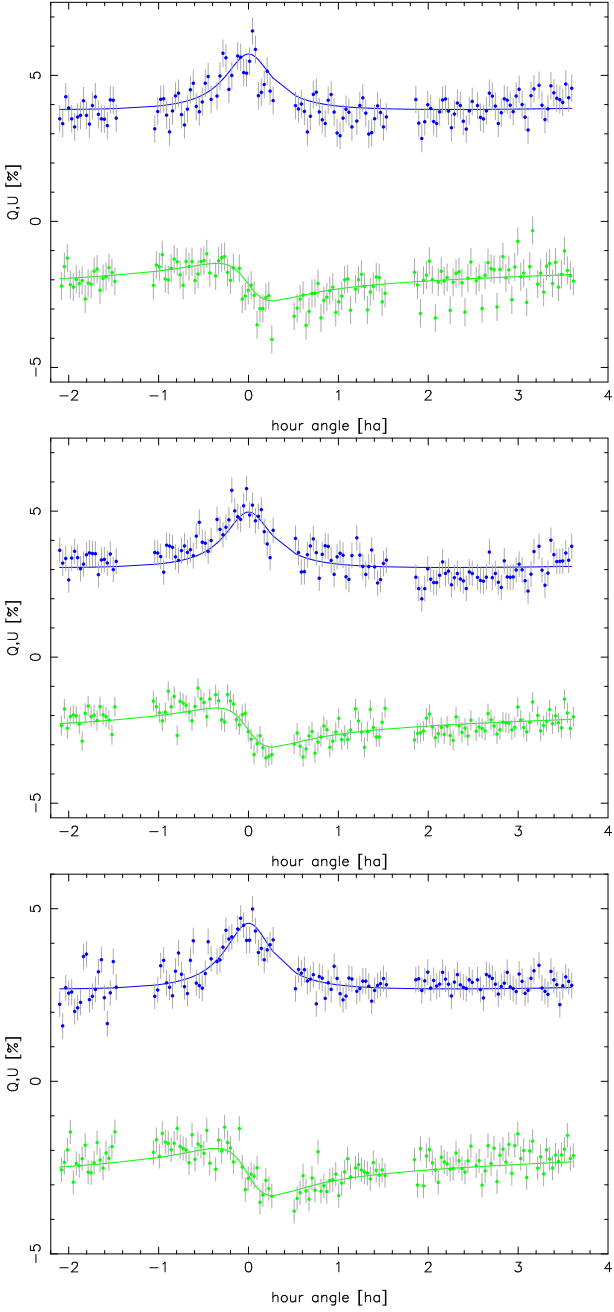

   \centering
   \includegraphics[angle=-90,width=8cm]{quha16NW.eps}
   \includegraphics[angle=-90,width=8cm]{quhaIRS16CC_2.eps}
   \includegraphics[angle=-90,width=8cm]{quhaIII.eps}
      \caption{Stokes-fits ($Q$ blue, $U$ green) for IRS16NW (upper panel), IRS16CC (middle), and S67 (lower panel). See Fig.~\ref{irsplots} and Table~\ref{polvar}.}
         \label{irsplots2}
\end{figure}

In Fig.~\ref{irsplots2} we show the confirmation of our model through Stokes-fits for IRS16NW, IRS16CC, and S67. In Fig.~\ref{lc3} we show two more Sgr~A* light curves analyzed with the more exact and elaborate method presented here and the 'boot strapping' method used before. The data used for this paper are summarized in Table~\ref{obs}.

\begin{figure}
   \centering
   \includegraphics[angle=-90,width=8cm]{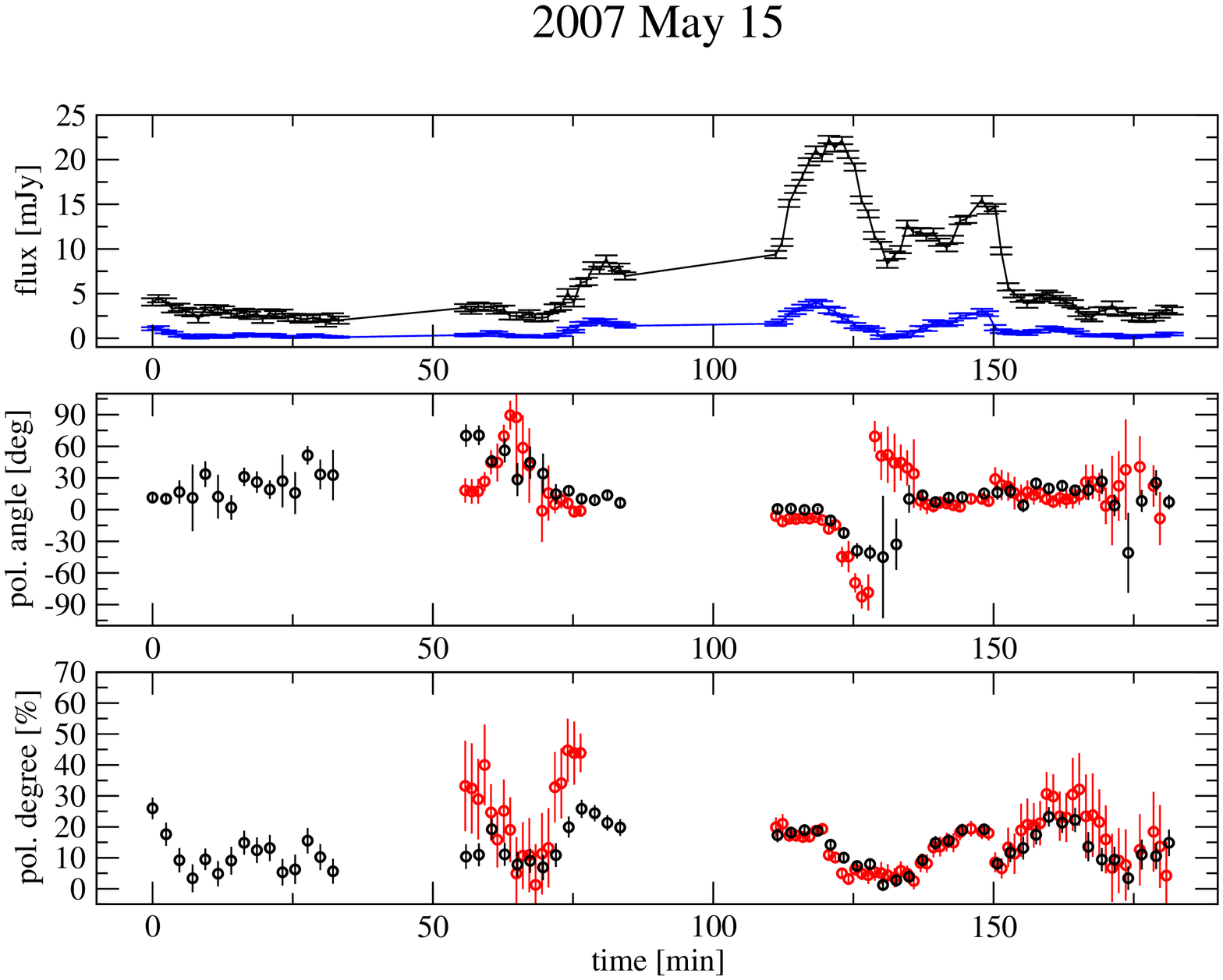}
      \caption{Polarimetric light curve from 2007 May 15. Plots analog to Fig.~\ref{lc1}. The meridian transit occurred at 86.5 min. }
         \label{lc2}
\end{figure}
\begin{figure}
   \centering
   \includegraphics[angle=-90,width=8cm]{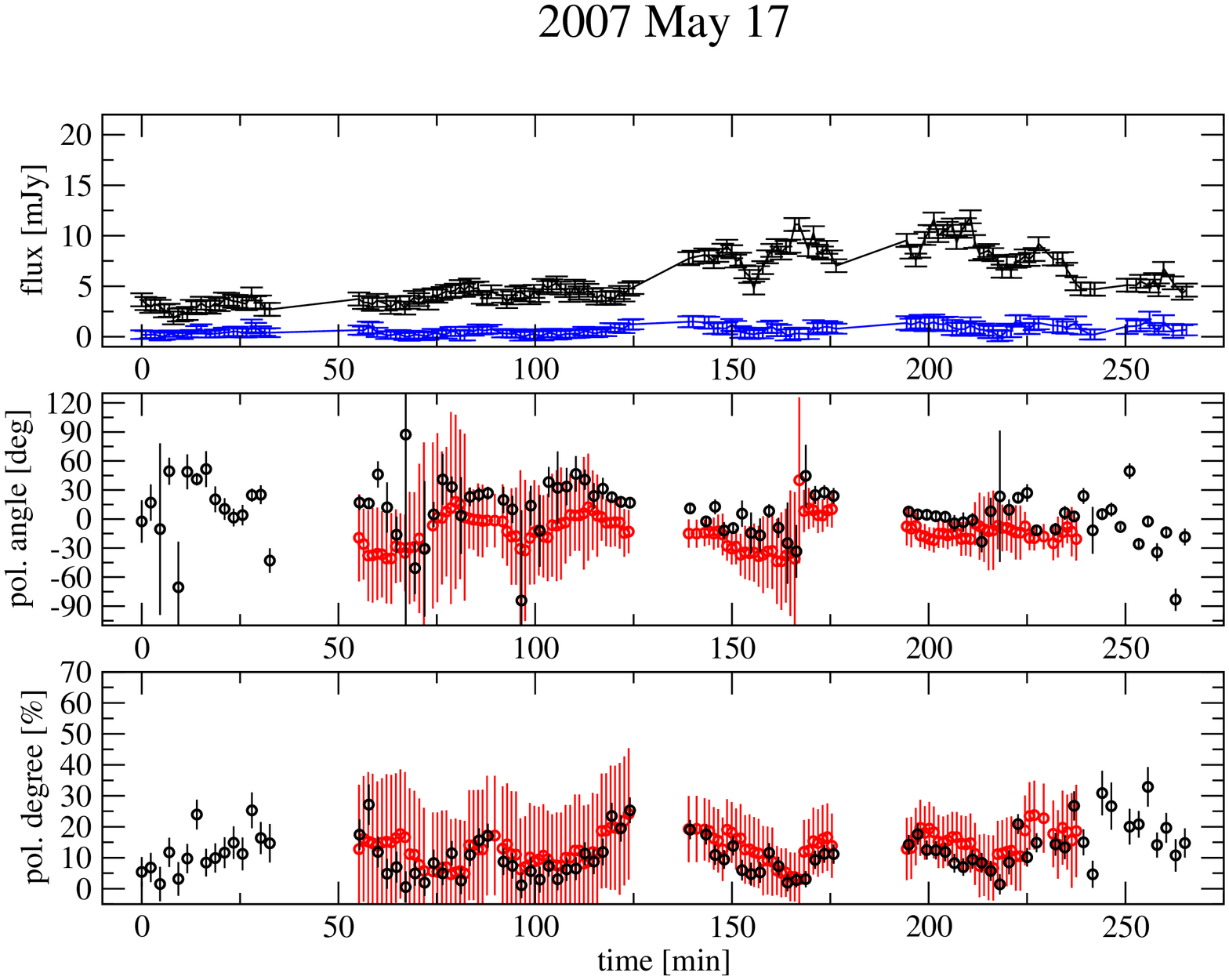}
      \caption{Polarimetric light curve from 2007 May 17. Plots analog to Fig.~\ref{lc1}. The meridian transit occurred at 126.4 min.}
         \label{lc3}
\end{figure}

\begin{figure}
   \centering
   \includegraphics[width=5cm]{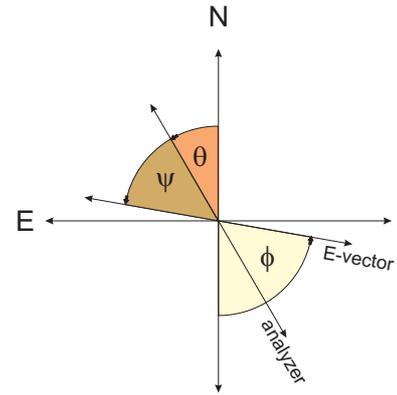}
      \caption{Orientations and angles for Eq.~\ref{eq2} and \ref{chan} with $\psi=\phi-\theta$. }
         \label{vect}
\end{figure}

\begin{table*}[b!]
\caption{Observations in the Ks-band.}
\label{obs}
\centering
\begin{tabular}{l l c l l l l }
\hline
\hline
\noalign{\smallskip}
Source  &  Date  & \# Frames & Time Interval (UT) & Project ID & PI & Setup\\
\noalign{\smallskip}
\hline
\noalign{\smallskip}
   GC \& Sgr~A* 
                & 2006 June 1  & 156 & 06:38:39 - 10:44:27 & 077.B-0552(A) & Eckart & Woll. \& HWP ($0^{\circ}/45^{\circ}$), S13\\
                & 2007 May 15  & 116 & 05:29:55 - 08:31:48 & 079.B-0084(A) & Eckart & Woll. \& HWP ($0^{\circ}/45^{\circ}$), S13\\
                & 2007 May 17  & 192 & 04:42:14 - 09:34:40 & 079.B-0084(A) & Eckart & Woll. \& HWP ($0^{\circ}/45^{\circ}$, S13)\\
  GC \& IRS16   & 2009 May 18  & 286 & 04:37:55 - 10:19:54 & 083.B-0031(A) & Eckart & Woll. \& HWP ($0^{\circ}/45^{\circ}$), S13\\
   IRS21        
                & 2005 July 30 & 18 & 06:44:50 - 07:01:32 & 075.B-0093(B) & Eckart &                    Woll. \& HWP ($0^{\circ}/30^{\circ}/60^{\circ}$), S13\\
   RCra88       & 2004 June 11 & 10 & 10:15:49 - 10:25:49 & 073.B-0084(A) & Genzel & Woll. \& HWP ($0^{\circ}/45^{\circ}$), S13\\
   WD1344       & 2007 April 08 / 09 & 10 &  & 079.D-0441(B) & Israel & Woll. \& rotator ($0^{\circ}/45^{\circ}$), S27\\
   WD2039-202   & 2007 August 18 / September 21 & 15 &  & 079.D-0444(A) & Israel & Woll. \& rotator ($0^{\circ}/45^{\circ}$), S27\\
   HD109055     & 2009 March 10 / 14 & 38 &  & 082.D-0137(B) & Israel & Woll. \& rotator ($0^{\circ}/45^{\circ}$), S27\\
\noalign{\smallskip}
\hline
\end{tabular}
\tablefoot{The column "setup" shows the facilities used and at which angles the data were obtained.}
\end{table*}

\end{appendix}

\end{document}